\documentstyle[12pt,aasms4]{article}

\def\PsfigVersion{1.10}
\def\setDriver{\DvipsDriver} 
\ifx\undefined\psfig\else\endinput\fi
\let\LaTeXAtSign=\@
\let\@=\relax
\edef\psfigRestoreAt{\catcode`\@=\number\catcode`@\relax}
\catcode`\@=11\relax
\newwrite\@unused
\def\ps@typeout#1{{\let\protect\string\immediate\write\@unused{#1}}}

\def\DvipsDriver{
	\ps@typeout{psfig/tex \PsfigVersion -dvips}
\def\PsfigSpecials{\DvipsSpecials} 	\def\ps@dir{/}
\def\ps@predir{} }
\def\OzTeXDriver{
	\ps@typeout{psfig/tex \PsfigVersion -oztex}
	\def\PsfigSpecials{\OzTeXSpecials}
	\def\ps@dir{:}
	\def\ps@predir{:}
	\catcode`\^^J=5
}

\def\figurepath{./:}

\def\DoPaths#1{\expandafter\EachPath#1\stoplist}
\def\leer{}
\def\EachPath#1:#2\stoplist{
  \ExistsFile{#1}{\SearchedFile}
  \ifx#2\leer
  \else
    \expandafter\EachPath#2\stoplist
  \fi}
\def\ps@dir{/}
\def\ExistsFile#1#2{%
   \openin1=\ps@predir#1\ps@dir#2
   \ifeof1
       \closein1
   \else
       \closein1
        \ifx\ps@founddir\leer
           \edef\ps@founddir{#1}
        \fi
   \fi}
\def\get@dir#1{%
  \def\ps@founddir{}
  \def\SearchedFile{#1}
  \DoPaths\figurepath
}

\def\@nnil{\@nil}
\def\@empty{}
\def\@psdonoop#1\@@#2#3{}
\def\@psdo#1:=#2\do#3{\edef\@psdotmp{#2}\ifx\@psdotmp\@empty \else
    \expandafter\@psdoloop#2,\@nil,\@nil\@@#1{#3}\fi}
\def\@psdoloop#1,#2,#3\@@#4#5{\def#4{#1}\ifx #4\@nnil \else
       #5\def#4{#2}\ifx #4\@nnil \else#5\@ipsdoloop #3\@@#4{#5}\fi\fi}
\def\@ipsdoloop#1,#2\@@#3#4{\def#3{#1}\ifx #3\@nnil 
       \let\@nextwhile=\@psdonoop \else
      #4\relax\let\@nextwhile=\@ipsdoloop\fi\@nextwhile#2\@@#3{#4}}
\def\@tpsdo#1:=#2\do#3{\xdef\@psdotmp{#2}\ifx\@psdotmp\@empty \else
    \@tpsdoloop#2\@nil\@nil\@@#1{#3}\fi}
\def\@tpsdoloop#1#2\@@#3#4{\def#3{#1}\ifx #3\@nnil 
       \let\@nextwhile=\@psdonoop \else
      #4\relax\let\@nextwhile=\@tpsdoloop\fi\@nextwhile#2\@@#3{#4}}
\ifx\undefined\fbox
\newdimen\fboxrule
\newdimen\fboxsep
\newdimen\ps@tempdima
\newbox\ps@tempboxa
\long\def\fbox#1{\leavevmode\setbox\ps@tempboxa\hbox{#1}\ps@tempdima\fboxrule
    \advance\ps@tempdima \fboxsep \advance\ps@tempdima \dp\ps@tempboxa
   \hbox{\lower \ps@tempdima\hbox
  {\vbox{\hrule height \fboxrule
          \hbox{\vrule width \fboxrule \hskip\fboxsep
          \vbox{\vskip\fboxsep \box\ps@tempboxa\vskip\fboxsep}\hskip 
                 \fboxsep\vrule width \fboxrule}
                 \hrule height \fboxrule}}}}
\fi
\newread\ps@stream
\newif\ifnot@eof       
\newif\if@noisy        
\newif\if@atend        
\newif\if@psfile       
{\catcode`\%=12\global\gdef\epsf@start{
\def\epsf@PS{PS}
\def\epsf@getbb#1{%
\openin\ps@stream=\ps@predir#1
\ifeof\ps@stream\ps@typeout{Error, File #1 not found}\else
   {\not@eoftrue \chardef\other=12
    \def\do##1{\catcode`##1=\other}\dospecials \catcode`\ =10
    \loop
       \if@psfile
	  \read\ps@stream to \epsf@fileline
       \else{
	  \obeyspaces
          \read\ps@stream to \epsf@tmp\global\let\epsf@fileline\epsf@tmp}
       \fi
       \ifeof\ps@stream\not@eoffalse\else
       \if@psfile\else
       \expandafter\epsf@test\epsf@fileline:. \\%
       \fi
          \expandafter\epsf@aux\epsf@fileline:. \\%
       \fi
   \ifnot@eof\repeat
   }\closein\ps@stream\fi}%
\long\def\epsf@test#1#2#3:#4\\{\def\epsf@testit{#1#2}
			\ifx\epsf@testit\epsf@start\else
\ps@typeout{Warning! File does not start with `\epsf@start'.  It may not be a PostScript file.}
			\fi
			\@psfiletrue} 
{\catcode`\%=12\global\let\epsf@percent=
\long\def\epsf@aux#1#2:#3\\{\ifx#1\epsf@percent
   \def\epsf@testit{#2}\ifx\epsf@testit\epsf@bblit
	\@atendfalse
        \epsf@atend #3 . \\%
	\if@atend	
	   \if@verbose{
		\ps@typeout{psfig: found `(atend)'; continuing search}
	   }\fi
        \else
        \epsf@grab #3 . . . \\%
        \not@eoffalse
        \global\no@bbfalse
        \fi
   \fi\fi}%
\def\epsf@grab #1 #2 #3 #4 #5\\{%
   \global\def\epsf@llx{#1}\ifx\epsf@llx\empty
      \epsf@grab #2 #3 #4 #5 .\\\else
   \global\def\epsf@lly{#2}%
   \global\def\epsf@urx{#3}\global\def\epsf@ury{#4}\fi}%
\def\epsf@atendlit{(atend)} 
\def\epsf@atend #1 #2 #3\\{%
   \def\epsf@tmp{#1}\ifx\epsf@tmp\empty
      \epsf@atend #2 #3 .\\\else
   \ifx\epsf@tmp\epsf@atendlit\@atendtrue\fi\fi}

\chardef\psletter = 11 
\chardef\other = 12

\newif \ifdebug 
\newif\ifc@mpute 
\c@mputetrue 

\let\then = \relax
\def\r@dian{pt }
\let\r@dians = \r@dian
\let\dimensionless@nit = \r@dian
\let\dimensionless@nits = \dimensionless@nit
\def\internal@nit{sp }
\let\internal@nits = \internal@nit
\newif\ifstillc@nverging
\def \Mess@ge #1{\ifdebug \then \message {#1} \fi}

{ 
	\catcode `\@ = \psletter
	\gdef \nodimen {\expandafter \n@dimen \the \dimen}
	\gdef \term #1 #2 #3%
	       {\edef \t@ {\the #1}
		\edef \t@@ {\expandafter \n@dimen \the #2\r@dian}%
		\t@rm {\t@} {\t@@} {#3}%
	       }
	\gdef \t@rm #1 #2 #3%
	       {{%
		\count 0 = 0
		\dimen 0 = 1 \dimensionless@nit
		\dimen 2 = #2\relax
		\Mess@ge {Calculating term #1 of \nodimen 2}%
		\loop
		\ifnum	\count 0 < #1
		\then	\advance \count 0 by 1
			\Mess@ge {Iteration \the \count 0 \space}%
			\Multiply \dimen 0 by {\dimen 2}%
			\Mess@ge {After multiplication, term = \nodimen 0}%
			\Divide \dimen 0 by {\count 0}%
			\Mess@ge {After division, term = \nodimen 0}%
		\repeat
		\Mess@ge {Final value for term #1 of 
				\nodimen 2 \space is \nodimen 0}%
		\xdef \Term {#3 = \nodimen 0 \r@dians}%
		\aftergroup \Term
	       }}
	\catcode `\p = \other
	\catcode `\t = \other
	\gdef \n@dimen #1pt{#1} 
}

\def \Divide #1by #2{\divide #1 by #2} 

\def \Multiply #1by #2
       {{
	\count 0 = #1\relax
	\count 2 = #2\relax
	\count 4 = 65536
	\Mess@ge {Before scaling, count 0 = \the \count 0 \space and
			count 2 = \the \count 2}%
	\ifnum	\count 0 > 32767 
	\then	\divide \count 0 by 4
		\divide \count 4 by 4
	\else	\ifnum	\count 0 < -32767
		\then	\divide \count 0 by 4
			\divide \count 4 by 4
		\else
		\fi
	\fi
	\ifnum	\count 2 > 32767 
	\then	\divide \count 2 by 4
		\divide \count 4 by 4
	\else	\ifnum	\count 2 < -32767
		\then	\divide \count 2 by 4
			\divide \count 4 by 4
		\else
		\fi
	\fi
	\multiply \count 0 by \count 2
	\divide \count 0 by \count 4
	\xdef \product {#1 = \the \count 0 \internal@nits}%
	\aftergroup \product
       }}

\def\r@duce{\ifdim\dimen0 > 90\r@dian \then   
		\multiply\dimen0 by -1
		\advance\dimen0 by 180\r@dian
		\r@duce
	    \else \ifdim\dimen0 < -90\r@dian \then  
		\advance\dimen0 by 360\r@dian
		\r@duce
		\fi
	    \fi}

\def\Sine#1%
       {{%
	\dimen 0 = #1 \r@dian
	\r@duce
	\ifdim\dimen0 = -90\r@dian \then
	   \dimen4 = -1\r@dian
	   \c@mputefalse
	\fi
	\ifdim\dimen0 = 90\r@dian \then
	   \dimen4 = 1\r@dian
	   \c@mputefalse
	\fi
	\ifdim\dimen0 = 0\r@dian \then
	   \dimen4 = 0\r@dian
	   \c@mputefalse
	\fi
	\ifc@mpute \then
		\divide\dimen0 by 180
		\dimen0=3.141592654\dimen0
		\dimen 2 = 3.1415926535897963\r@dian 
		\divide\dimen 2 by 2 
		\Mess@ge {Sin: calculating Sin of \nodimen 0}%
		\count 0 = 1 
		\dimen 2 = 1 \r@dian 
		\dimen 4 = 0 \r@dian 
		\loop
			\ifnum	\dimen 2 = 0 
			\then	\stillc@nvergingfalse 
			\else	\stillc@nvergingtrue
			\fi
			\ifstillc@nverging 
			\then	\term {\count 0} {\dimen 0} {\dimen 2}%
				\advance \count 0 by 2
				\count 2 = \count 0
				\divide \count 2 by 2
				\ifodd	\count 2 
				\then	\advance \dimen 4 by \dimen 2
				\else	\advance \dimen 4 by -\dimen 2
				\fi
		\repeat
	\fi		
			\xdef \sine {\nodimen 4}%
       }}

\def\Cosine#1{\ifx\sine\UnDefined\edef\Savesine{\relax}\else
		             \edef\Savesine{\sine}\fi
	{\dimen0=#1\r@dian\advance\dimen0 by 90\r@dian
	 \Sine{\nodimen 0}
	 \xdef\cosine{\sine}
	 \xdef\sine{\Savesine}}}	      

\def\psdraft{
	\def\@psdraft{0}
}
\def\psfull{
	\def\@psdraft{100}
}

\psfull

\newif\if@scalefirst
\def\psscalefirst{\@scalefirsttrue}
\def\psrotatefirst{\@scalefirstfalse}
\psrotatefirst

\newif\if@draftbox
\def\psnodraftbox{
	\@draftboxfalse
}
\def\psdraftbox{
	\@draftboxtrue
}
\@draftboxtrue

\newif\if@prologfile
\newif\if@postlogfile
\def\pssilent{
	\@noisyfalse
}
\def\psnoisy{
	\@noisytrue
}
\psnoisy
\newif\if@bbllx
\newif\if@bblly
\newif\if@bburx
\newif\if@bbury
\newif\if@height
\newif\if@width
\newif\if@rheight
\newif\if@rwidth
\newif\if@angle
\newif\if@clip
\newif\if@verbose
\def\@p@@sclip#1{\@cliptrue}
\newif\if@decmpr
\def\@p@@sfigure#1{\def\@p@sfile{null}\def\@p@sbbfile{null}\@decmprfalse
   \openin1=\ps@predir#1
   \ifeof1
	\closein1
	\get@dir{#1}
	\ifx\ps@founddir\leer
		\openin1=\ps@predir#1.bb
		\ifeof1
			\closein1
			\get@dir{#1.bb}
			\ifx\ps@founddir\leer
				\ps@typeout{Can't find #1 in \figurepath}
			\else
				\@decmprtrue
				\def\@p@sfile{\ps@founddir\ps@dir#1}
				\def\@p@sbbfile{\ps@founddir\ps@dir#1.bb}
			\fi
		\else
			\closein1
			\@decmprtrue
			\def\@p@sfile{#1}
			\def\@p@sbbfile{#1.bb}
		\fi
	\else
		\def\@p@sfile{\ps@founddir\ps@dir#1}
		\def\@p@sbbfile{\ps@founddir\ps@dir#1}
	\fi
   \else
	\closein1
	\def\@p@sfile{#1}
	\def\@p@sbbfile{#1}
   \fi
}
\def\@p@@sfile#1{\@p@@sfigure{#1}}
\def\@p@@sbbllx#1{
		\@bbllxtrue
		\dimen100=#1
		\edef\@p@sbbllx{\number\dimen100}
}
\def\@p@@sbblly#1{
		\@bbllytrue
		\dimen100=#1
		\edef\@p@sbblly{\number\dimen100}
}
\def\@p@@sbburx#1{
		\@bburxtrue
		\dimen100=#1
		\edef\@p@sbburx{\number\dimen100}
}
\def\@p@@sbbury#1{
		\@bburytrue
		\dimen100=#1
		\edef\@p@sbbury{\number\dimen100}
}
\def\@p@@sheight#1{
		\@heighttrue
		\dimen100=#1
   		\edef\@p@sheight{\number\dimen100}
}
\def\@p@@swidth#1{
		\@widthtrue
		\dimen100=#1
		\edef\@p@swidth{\number\dimen100}
}
\def\@p@@srheight#1{
		\@rheighttrue
		\dimen100=#1
		\edef\@p@srheight{\number\dimen100}
}
\def\@p@@srwidth#1{
		\@rwidthtrue
		\dimen100=#1
		\edef\@p@srwidth{\number\dimen100}
}
\def\@p@@sangle#1{
		\@angletrue
		\edef\@p@sangle{#1} 
}
\def\@p@@ssilent#1{ 
		\@verbosefalse
}
\def\@p@@sprolog#1{\@prologfiletrue\def\@prologfileval{#1}}
\def\@p@@spostlog#1{\@postlogfiletrue\def\@postlogfileval{#1}}
\def\@cs@name#1{\csname #1\endcsname}
\def\@setparms#1=#2,{\@cs@name{@p@@s#1}{#2}}
\def\ps@init@parms{
		\@bbllxfalse \@bbllyfalse
		\@bburxfalse \@bburyfalse
		\@heightfalse \@widthfalse
		\@rheightfalse \@rwidthfalse
		\def\@p@sbbllx{}\def\@p@sbblly{}
		\def\@p@sbburx{}\def\@p@sbbury{}
		\def\@p@sheight{}\def\@p@swidth{}
		\def\@p@srheight{}\def\@p@srwidth{}
		\def\@p@sangle{0}
		\def\@p@sfile{} \def\@p@sbbfile{}
		\def\@p@scost{10}
		\def\@sc{}
		\@prologfilefalse
		\@postlogfilefalse
		\@clipfalse
		\if@noisy
			\@verbosetrue
		\else
			\@verbosefalse
		\fi
}
\def\parse@ps@parms#1{
	 	\@psdo\@psfiga:=#1\do
		   {\expandafter\@setparms\@psfiga,}}
\newif\ifno@bb
\def\bb@missing{
	\if@verbose{
		\ps@typeout{psfig: searching \@p@sbbfile \space  for bounding box}
	}\fi
	\no@bbtrue
	\epsf@getbb{\@p@sbbfile}
        \ifno@bb \else \bb@cull\epsf@llx\epsf@lly\epsf@urx\epsf@ury\fi
}	
\def\bb@cull#1#2#3#4{
	\dimen100=#1 bp\edef\@p@sbbllx{\number\dimen100}
	\dimen100=#2 bp\edef\@p@sbblly{\number\dimen100}
	\dimen100=#3 bp\edef\@p@sbburx{\number\dimen100}
	\dimen100=#4 bp\edef\@p@sbbury{\number\dimen100}
	\no@bbfalse
}
\newdimen\p@intvaluex
\newdimen\p@intvaluey
\def\rotate@#1#2{{\dimen0=#1 sp\dimen1=#2 sp
		  \global\p@intvaluex=\cosine\dimen0
		  \dimen3=\sine\dimen1
		  \global\advance\p@intvaluex by -\dimen3
		  \global\p@intvaluey=\sine\dimen0
		  \dimen3=\cosine\dimen1
		  \global\advance\p@intvaluey by \dimen3
		  }}
\def\compute@bb{
		\no@bbfalse
		\if@bbllx \else \no@bbtrue \fi
		\if@bblly \else \no@bbtrue \fi
		\if@bburx \else \no@bbtrue \fi
		\if@bbury \else \no@bbtrue \fi
		\ifno@bb \bb@missing \fi
		\ifno@bb \ps@typeout{FATAL ERROR: no bb supplied or found}
			\no-bb-error
		\fi
		\count203=\@p@sbburx
		\count204=\@p@sbbury
		\advance\count203 by -\@p@sbbllx
		\advance\count204 by -\@p@sbblly
		\edef\ps@bbw{\number\count203}
		\edef\ps@bbh{\number\count204}
		\if@angle 
			\Sine{\@p@sangle}\Cosine{\@p@sangle}
	        	{\dimen100=\maxdimen\xdef\r@p@sbbllx{\number\dimen100}
					    \xdef\r@p@sbblly{\number\dimen100}
			                    \xdef\r@p@sbburx{-\number\dimen100}
					    \xdef\r@p@sbbury{-\number\dimen100}}
                        \def\minmaxtest{
			   \ifnum\number\p@intvaluex<\r@p@sbbllx
			      \xdef\r@p@sbbllx{\number\p@intvaluex}\fi
			   \ifnum\number\p@intvaluex>\r@p@sbburx
			      \xdef\r@p@sbburx{\number\p@intvaluex}\fi
			   \ifnum\number\p@intvaluey<\r@p@sbblly
			      \xdef\r@p@sbblly{\number\p@intvaluey}\fi
			   \ifnum\number\p@intvaluey>\r@p@sbbury
			      \xdef\r@p@sbbury{\number\p@intvaluey}\fi
			   }
			\rotate@{\@p@sbbllx}{\@p@sbblly}
			\minmaxtest
			\rotate@{\@p@sbbllx}{\@p@sbbury}
			\minmaxtest
			\rotate@{\@p@sbburx}{\@p@sbblly}
			\minmaxtest
			\rotate@{\@p@sbburx}{\@p@sbbury}
			\minmaxtest
			\edef\@p@sbbllx{\r@p@sbbllx}\edef\@p@sbblly{\r@p@sbblly}
			\edef\@p@sbburx{\r@p@sbburx}\edef\@p@sbbury{\r@p@sbbury}
		\fi
		\count203=\@p@sbburx
		\count204=\@p@sbbury
		\advance\count203 by -\@p@sbbllx
		\advance\count204 by -\@p@sbblly
		\edef\@bbw{\number\count203}
		\edef\@bbh{\number\count204}
}
\def\in@hundreds#1#2#3{\count240=#2 \count241=#3
		     \count100=\count240	
		     \divide\count100 by \count241
		     \count101=\count100
		     \multiply\count101 by \count241
		     \advance\count240 by -\count101
		     \multiply\count240 by 10
		     \count101=\count240	
		     \divide\count101 by \count241
		     \count102=\count101
		     \multiply\count102 by \count241
		     \advance\count240 by -\count102
		     \multiply\count240 by 10
		     \count102=\count240	
		     \divide\count102 by \count241
		     \count200=#1\count205=0
		     \count201=\count200
			\multiply\count201 by \count100
		 	\advance\count205 by \count201
		     \count201=\count200
			\divide\count201 by 10
			\multiply\count201 by \count101
			\advance\count205 by \count201
		     \count201=\count200
			\divide\count201 by 100
			\multiply\count201 by \count102
			\advance\count205 by \count201
		     \edef\@result{\number\count205}
}
\def\compute@wfromh{
		\in@hundreds{\@p@sheight}{\@bbw}{\@bbh}
		\edef\@p@swidth{\@result}
}
\def\compute@hfromw{
	        \in@hundreds{\@p@swidth}{\@bbh}{\@bbw}
		\edef\@p@sheight{\@result}
}
\def\compute@handw{
		\if@height 
			\if@width
			\else
				\compute@wfromh
			\fi
		\else 
			\if@width
				\compute@hfromw
			\else
				\edef\@p@sheight{\@bbh}
				\edef\@p@swidth{\@bbw}
			\fi
		\fi
}
\def\compute@resv{
		\if@rheight \else \edef\@p@srheight{\@p@sheight} \fi
		\if@rwidth \else \edef\@p@srwidth{\@p@swidth} \fi
}
\def\compute@sizes{
	\compute@bb
	\if@scalefirst\if@angle
	\if@width
	   \in@hundreds{\@p@swidth}{\@bbw}{\ps@bbw}
	   \edef\@p@swidth{\@result}
	\fi
	\if@height
	   \in@hundreds{\@p@sheight}{\@bbh}{\ps@bbh}
	   \edef\@p@sheight{\@result}
	\fi
	\fi\fi
	\compute@handw
	\compute@resv}
\def\OzTeXSpecials{
	\special{empty.ps /@isp {true} def}
	\special{empty.ps \@p@swidth \space \@p@sheight \space
			\@p@sbbllx \space \@p@sbblly \space
			\@p@sbburx \space \@p@sbbury \space
			startTexFig \space }
	\if@clip{
		\if@verbose{
			\ps@typeout{(clip)}
		}\fi
		\special{empty.ps doclip \space }
	}\fi
	\if@angle{
		\if@verbose{
			\ps@typeout{(rotate)}
		}\fi
		\special {empty.ps \@p@sangle \space rotate \space} 
	}\fi
	\if@prologfile
	    \special{\@prologfileval \space } \fi
	\if@decmpr{
		\if@verbose{
			\ps@typeout{psfig: Compression not available
			in OzTeX version \space }
		}\fi
	}\else{
		\if@verbose{
			\ps@typeout{psfig: including \@p@sfile \space }
		}\fi
		\special{epsf=\ps@predir\@p@sfile \space }
	}\fi
	\if@postlogfile
	    \special{\@postlogfileval \space } \fi
	\special{empty.ps /@isp {false} def}
}
\def\DvipsSpecials{
	\special{ps::[begin] 	\@p@swidth \space \@p@sheight \space
			\@p@sbbllx \space \@p@sbblly \space
			\@p@sbburx \space \@p@sbbury \space
			startTexFig \space }
	\if@clip{
		\if@verbose{
			\ps@typeout{(clip)}
		}\fi
		\special{ps:: doclip \space }
	}\fi
	\if@angle
		\if@verbose{
			\ps@typeout{(clip)}
		}\fi
		\special {ps:: \@p@sangle \space rotate \space} 
	\fi
	\if@prologfile
	    \special{ps: plotfile \@prologfileval \space } \fi
	\if@decmpr{
		\if@verbose{
			\ps@typeout{psfig: including \@p@sfile.Z \space }
		}\fi
		\special{ps: plotfile "`zcat \@p@sfile.Z" \space }
	}\else{
		\if@verbose{
			\ps@typeout{psfig: including \@p@sfile \space }
		}\fi
		\special{ps: plotfile \@p@sfile \space }
	}\fi
	\if@postlogfile
	    \special{ps: plotfile \@postlogfileval \space } \fi
	\special{ps::[end] endTexFig \space }
}
\def\psfig#1{\vbox {
	\ps@init@parms
	\parse@ps@parms{#1}
	\compute@sizes
	\ifnum\@p@scost<\@psdraft{
		\PsfigSpecials 
		\vbox to \@p@srheight sp{
			\hbox to \@p@srwidth sp{
				\hss
			}
		\vss
		}
	}\else{
		\if@draftbox{		
			\hbox{\fbox{\vbox to \@p@srheight sp{
			\vss
			\hbox to \@p@srwidth sp{ \hss 
			 \hss }
			\vss
			}}}
		}\else{
			\vbox to \@p@srheight sp{
			\vss
			\hbox to \@p@srwidth sp{\hss}
			\vss
			}
		}\fi

	}\fi
}}
\psfigRestoreAt
\setDriver
\let\@=\LaTeXAtSign

\def\sun{$_\odot$}
\def\lapp{\lower2pt\hbox{$\buildrel {\scriptstyle <}
   \over {\scriptstyle\sim}$}}
\def\gapp{\lower2pt\hbox{$\buildrel {\scriptstyle >}
   \over {\scriptstyle\sim}$}}
\def\deriv#1#2{{{\rm d}#1\over {\rm d}#2}}
\def\msun{{\rm M}_{\odot}}
\def\rsun{{\rm R}_{\odot}}
\def\lsun{{\rm L}_{\odot}}
\def\mdot{\dot \rm M}
\def\mpy{{\rm M}_{\odot} {\rm yr}^{-1}} 

\newcommand{\be}{\begin{equation}}
\newcommand{\ee}{\end{equation}}
\newcommand{\bcen}{\begin{center}}
\newcommand{\ecen}{\end{center}}
\newcommand{\di}{\partial}
\newcommand{\bprime}{\mbox{$\beta^{\prime}$}}
\newcommand{\subsect}{\subsection}
\newcommand{\sect}{\section}
\newcommand{\Ang}{\mbox{\raisebox{1.7ex}{$\tiny\circ$}\hspace{-0.5em}A}}
\newcommand{\kms}{\mbox{ km~s$^{-1}$}}
\newcommand{\Mpc}{\mbox{M$_\odot$~pc$^{-3}$}}
\newcommand{\msolar}{\mbox{M$_\odot$}}
\newcommand{\lsolar}{\mbox{L$_\odot$}}
\newcommand{\kpc}{\mbox{kpc}}

\def\deg{\ifmmode ^{\circ}
         \else $^{\circ}$\fi}
\def\pdeg{\ifmmode $\setbox0=\hbox{$^{\circ}$}\rlap{\hskip.11\wd0 .}$^{\circ}
          \else \setbox0=\hbox{$^{\circ}$}\rlap{\hskip.11\wd0 .}$^{\circ}$\fi}
\def\arcs{\ifmmode {^{\scriptscriptstyle\prime\prime}}
          \else $^{\scriptscriptstyle\prime\prime}$\fi}
\def\arcm{\ifmmode {^{\scriptscriptstyle\prime}}
          \else $^{\scriptscriptstyle\prime}$\fi}
\newdimen\sa  \newdimen\sb
\def\parcs{\sa=.07em \sb=.03em
     \ifmmode $\rlap{.}$^{\scriptscriptstyle\prime\kern -\sb\prime}$\kern -\sa$
     \else \rlap{.}$^{\scriptscriptstyle\prime\kern -\sb\prime}$\kern -\sa\fi}
\def\parcm{\sa=.08em \sb=.03em
     \ifmmode $\rlap{.}\kern\sa$^{\scriptscriptstyle\prime}$\kern-\sb$
     \else \rlap{.}\kern\sa$^{\scriptscriptstyle\prime}$\kern-\sb\fi}
\def\gtorder{\mathrel{\raise.3ex\hbox{$>$}\mkern-14mu
             \lower0.6ex\hbox{$\sim$}}}
\def\ltorder{\mathrel{\raise.3ex\hbox{$<$}\mkern-14mu
             \lower0.6ex\hbox{$\sim$}}}

  \def\aa{{ A\&A}}
  \def\aj{{ AJ}}
  \def\annrev{{ ARA\&A}}
  \def\apj{{ ApJ}}
  \def\apjl{{ ApJ}}
  \def\apjs{{ ApJS}}
  \def\mn{{ MNRAS}}
  \def\Nature{{ Nature}}
  \def\science{{ Science}}
  \def\pasp{{ PASP}}
  \def\rmp{{ RevModPhys}}

\def\refindent{\par\penalty-100\noindent\parskip=4pt plus1pt
               \hangindent=3pc\hangafter=1\null}
\def\ref#1#2#3#4{\refindent#2, {#1\/,\ }{#3}, #4}
\def\reft#1#2#3#4#5#6{\refindent#2 {\it #6}, #3, {\it #1\/,\ }{\bf#4}, #5.}
\def\reftb#1#2#3#4#5#6{\refindent#2 #3, {\it #6}, {\it #1\/,\ }{\bf#4}, #5.}
\def\refbook#1{\refindent#1}
\def\preprint#1#2#3{\refindent#1, #2, {\it #3 preprint} }
\def\preprintt#1#2#3#4{\refindent#1, #2 {\it #4}, {\it #3 preprint}.}
\def\refinpress#1#2{\refindent#1, {\it #2, in press}.}
\def\reftinpress#1#2#3#4{\refindent#1 {#4}, #2, {\it #3, in press}}
\def\reftsubmit#1#2#3#4{\refindent#1 {#4}, #2, {\it #3, submitted}}
\def\bysame{\hbox to 50pt{\leaders\hrule height 2.4pt depth -2pt\hfill .\ }}

\def\sdml{$\sigma_{DM}$-$L$ }
\def\sdmlmu{$\sigma_{DM}$-$L$-$\mu$ }
\def\sdmlre{$\sigma_{DM}$-$L$-$R_e$ }

\begin{document}

\title{Rebuilding the Cepheid Distance Scale I: \\
   A Global Analysis of Cepheid Mean Magnitudes }
\author{C.S. Kochanek}
\affil{Harvard-Smithsonian Center for Astrophysics, MS-51\protect \\
       60 Garden Street \protect \\
       Cambridge MA 02138 }
\authoremail{ckochanek@cfa.harvard.edu}

\begin{abstract}
We develop a statistical method for using multicolor photometry to determine 
distances using Cepheid variables including the effects of temperature, extinction, 
and metallicity and apply it to UBVRIJHK photometry of 694 Cepheids in 17 galaxies. 
We derive homogeneous distance, extinction and uncertainty estimates for four
models, starting from the standard extragalactic method and then adding the
physical effects of temperature distributions, extinction distributions, requiring
positive definite extinctions, and metallicity.  While we find general agreement 
with published distances when we make similar systematic assumptions, 
there is a clear problem in the standard distances because they require  
Cepheids with negative extinctions, particularly in low
metallicity galaxies, unless the mean LMC extinction exceeds $E(B-V)\gtorder 0.25$.   
The problem can be explained by the physically expected  metallicity dependence 
of the Cepheid distance scale, where metal-poor Cepheids are hotter and fainter 
than metal-rich Cepheids.  For V and I we found that the mean magnitude change
is $-0.14\pm0.14$ mag/dex and the mean color change is $0.13\pm0.04$ mag/dex,
with the change in color dominating the change in distance.  The effect on
Type Ia supernova estimates of the Hubble constant is dramatic because most
were found in the metal poor galaxies with the bluest Cepheids. The 
Type Ia Multi-color Light Curve Shape (MLCS) method estimate for $H_0$ formally 
rises from $69\pm8\kms$ Mpc$^{-1}$ to $80\pm6 \kms$ Mpc$^{-1}$ with the metallicity correction.
\end{abstract}


\section{Introduction}

Cepheid variables are fundamental to most extragalactic distance
estimates, determinations of the Hubble constant, and models for
the structure of the Galaxy because the Cepheid period-luminosity
(PL) relations are accepted as one of the most accurate primary
distance indicators.  In the last few years, the number
of extragalactic Cepheids has exploded due to 
the two large extragalactic surveys using the Hubble Space Telescope
(the Extragalactic Distance Scale Key Project (Freedman et al. 1994,
Ferrarese et al. 1996, Graham et al. 1996, Kelson et al. 1996, 
Silbermann et al. 1996) and the Type Ia Supernova Calibration Project 
(Saha et al. 1994, 1995, 1996ab, 1997)), and the microlensing surveys of the Magellanic 
Clouds (e.g. Beaulieu et al. 1996, Welch et al. 1997).
The goal of the HST projects is to determine the Hubble
constant with 10\% (0.2 mag) accuracy, which requires uncertainties 
in the Cepheid distance estimates, including all systematic uncertainties, 
that are still smaller.  In order to reach these goals, the 
observational projects (see the review by Freedman 1997) have
expanded the range of filters used to study Cepheids, particularly into
the infrared, introduced systematic corrections for extinction,
and tried to find good empirical tests for the effects of metallicity
on Cepheid distance estimates.  

The physical basis for Cepheids as distance estimators rests on 
two foundations (see the reviews by Feast \& Walker (1987),
Madore \& Freedman (1991), and Tanvir (1996)).
The first foundation is stellar structure, which closely correlates
the mass, radius, luminosity, temperature, and oscillation period
of a star, so that in any localized region of the H-R diagram there is 
period-luminosity-color (``PLC'') relation stating that the luminosity 
can be determined from the color (temperature) and the period (mass and radius)
of oscillation.   The second foundation is that
the physics of the oscillations limits the Cepheids to a narrow
range of luminosity and temperature called the instability strip.  As a result,
any projection of the three-dimensional PLC space onto a two-dimensional
subspace produces a tightly correlated relation.  In particular, the
period-luminosity (PL) correlations are defined by projecting over
the temperature/color distribution.  Both the PLC relations and the location 
of the instability strip are expected to be functions of composition (e.g.
Stothers 1988, Stift 1990, Chiosi, Wood \& Capitanio 1993), 
although there is great debate about the magnitude of the dependence 
and its measurement (e.g. Caldwell \& Coulson (1986, 1987), Freedman \& Madore (1990)
Gould (1994), Stift (1995), Sasselov et al. (1996)).  The sense, however, is that metal
rich Cepheids are cooler and brighter than metal poor Cepheids at fixed period.

The goal of any analysis of Cepheid data is to accurately determine the 
distance to the Cepheid and its uncertainty, after compensating for 
the effects of period, temperature, composition, and extinction.  
Existing Cepheid analysis methods are divided into extragalactic 
and Galactic approaches. 
For extragalactic systems the analysis must determine the common 
distance of an ensemble of Cepheids, usually based on two-color,
low-accuracy photometry, with poor phase coverage.  The standard
approach (see Madore \& Freedman 1991) uses only the correlation between 
luminosity and period (the PL relations) to estimate the distance modulus and the mean
extinction.  Gould (1994) pointed out that the method is statistically
inefficient because it ignores the strong correlations in the
residuals (i.e. the PLC relations).   Galactic analyses (e.g. Caldwell \&
Coulson 1986, 1987, Caldwell \& Laney 1991, Laney \& Stobie 1993, 1994) must estimate 
the distances and extinctions of individual Cepheids using 
three-color photometry and PLC relations because the Cepheids no
longer lie at a common distance. 
The standard extragalactic approach explicitly ignores the PLC relations,
and the standard Galactic approach (in some senses) ignores the instability
strip.  

In the following sections we develop a self-consistent physical and
mathematical analysis of multi-color Cepheid mean magnitudes as the
first in a series papers reanalyzing the Cepheid distance scale.  
Our approach differs from the standard approaches in three major ways.  The
first difference is that we analyze all the data simultaneously rather than
one galaxy at a time.  The use of Cepheids as distance indicators
is predicated on the homogeneity of their physical properties, so 
either all the data can be self-consistently and simultaneously
analyzed or we must reject the Cepheids as distance indicators.   
Moreover, the distance estimates (and other physical parameters)
are very highly correlated.  Any change in the distance or extinction estimate
for one Cepheid galaxy requires a simultaneous change in the values for
all Cepheid galaxies. The second difference is that we avoid
the false dichotomy between the Galactic and extragalactic analysis
procedures, and demonstrate how to reconcile the two approaches.
The resulting statistical method has greater
statistical efficiency than traditional methods, can model all measured 
colors simultaneously, and allows for better control, treatment, and understanding of 
systematic problems such as individual Cepheid extinctions and the
effects of metallicity.  The resulting scheme, although based on
a physical model, closely resembles
the empirical treatments of Gould (1994) and Sasselov et al. (1996).
The third difference is that we explore the important systematic errors that 
can affect the Cepheid distance estimates, particularly the physics of extinction, 
temperature and metallicity.
While treatments of extragalactic Cepheid distances generally recognize the
existence of these systematic uncertainties, they are rarely included in the
distance estimates or their uncertainties.
We develop our analysis method in \S2, and compare it to the existing
techniques. In \S3 we examine the Cepheid correlations (PL, PLC relations
etc.) and the resulting distance estimates.  We summarize our results, their
shortcomings, and possible solutions in \S4.

\section{Cepheid Distance Estimates}

We assume that the bolometric luminosity and the bolometric correction
are linear functions of the period $p=\log P/P_0$, the effective temperature $t=\log T_e/T_0$, 
and the (logarithmic) metallicity $Z$, so that the (intensity mean) apparent magnitude 
$V_k$ of a Cepheid in band $k$, at distance modulus $\mu$, with extinction $E$ is 
\begin{equation}
  V_k = V_{0k} + \alpha_k p + \beta_k t + \gamma_k Z + \mu d_k + E R_k 
\end{equation}
where $V_{0k}$ is the magnitude zero-point vector,
$R_k$ is the reddening vector, $d_k \equiv 1$ is the distance vector,
and $\log P_0=1.4$ and $T_0$ are reference periods and temperatures.  
We assume that the slopes $\alpha_k$ and $\beta_k$ are independent of metallicity,
although theory predicts a weak dependence (see Stothers 1988, Stift 1990, 1995, Chiosi et al. 1993). 
If we neglect the dependence of 
bolometric corrections on surface gravity, $\alpha_k=\alpha_j$.  We call 
eqn. (1) a period-luminosity-color (PLC) relation, although standard
PLC relations (see Feast \& Walker 1987) replace the effective temperature by a color.  
We neglect additional variables such as the Cepheid's age, the instability strip 
crossing number, the helium abundance, and non-linearities in the PLC relation
(see Caldwell \& Coulson 1986), so we assume that 
the relations defined by eqn. (1) have an intrinsic width of $\sigma_{PLCk}$.

Formally, if we know the precise values of the vectors $\alpha_k$,
$\beta_k$, $\gamma_k$ and $R_k$, the vectors are not degenerate with $d_k$
($\equiv$ distance), and we possess accurate photometry
of the Cepheid in a sufficiently large number of bands, then the PLC
relations can be used to determine the distance to a particular Cepheid.
Unfortunately, since we must determine $\alpha_k$, $\beta_k$ and $\gamma_k$
as we proceed from a small number of colors, it is difficult to use the 
PLC relations to determine distances without adding additional constraints.  
The additional constraints used in standard Cepheid analyses (e.g. Madore \& 
Freedman 1991) correspond to adding priors on the distances and extinctions.
We now develop a general mathematical description for fitting Cepheid 
magnitudes that includes the standard methods as subcases or limits of a 
more general model.

The PLC relations contain no information about which stars pulsate, and 
the most important prior information is the location and width of the 
instability strip.  The standard extragalactic method uses the instability
strip by averaging the distribution over temperature and using only
the PL relations to determine distances.  We parametrize the instability strip by
a period-color (PC) relation defining the temperature distribution
for stars at a fixed period using a likelihood function
\begin{equation}
     L(t | p ) \propto 
  \exp\left[ - { (t - \delta_Z Z - \delta_P p )^2 
     \over 2\sigma_{PC}^2 } \right]
\end{equation}
including a possible shift in the location of the instability strip with
metallicity (e.g. Stift 1990, Chiosi et al. 1993).  We assume that the width of the instability strip
is constant, although the observed narrowing of the strip at short
periods (e.g. Fernie 1990) could be modeled by making $\sigma_{PC}$ 
a function of period (e.g. Gould 1994).  We chose to parametrize the 
instability strip with a PC 
relation rather than a luminosity-color relation (i.e. the H-R diagram)
because it is distance independent and uses the well-defined period
as the independent variable instead of the luminosity.  
From these two assumptions we can derive all the standard relations used 
to study Cepheids and their correlations.  For example,
the PL relation in band $V_k$ is 
\begin{equation}
   { \int V_k L(t |p) dt  \over 
     \int L(t|p) d t } 
     = \langle V_k \rangle = V_{0k} + \alpha_k' p + 
    \gamma_k' \langle Z\rangle  + \mu d_k +
      \langle E\rangle R_k
\end{equation}
where $\alpha_k' = \alpha_k + \beta_k \delta_P$ and 
$\gamma_k' = \gamma_k + \beta_k \delta_Z$, and the dispersion in the relation is
$\sigma_{PLCk}^2+\beta_k^2 \sigma_{PC}^2\simeq \beta_k^2\sigma_{PC}^2$
since $\sigma_{PLCk} \ll \sigma_{PC}$.  From here on we use the deviation
of the temperature from the expected mean, $\delta t = t - \delta_Z Z - \delta_P p$,
which also shifts the period and metallicity vectors to the $\alpha_k'$ and $\gamma_k'$
appearing in the PL relation (3).  We cannot independently determine $\alpha_k$, $\gamma_k$,
$\delta_Z$ and $\delta_P$ without absolute temperature references, so we restrict our
solutions to determinations of $\delta t$, $\alpha'_k$, and $\gamma'_k$.    
We measure distances, 
extinctions, and metallicities relative to the LMC values of $\langle \mu \rangle_{LMC}$,
$\langle E\rangle_{LMC}$, and $Z_{LMC}\equiv [O/H]_{LMC}$.  
We use the oxygen abundance for the metallicity variable because it is the
only abundance available for most of the extragalactic systems.
We can also constrain the models using prior information on the distance modulus, 
extinction, and metallicity.  In external galaxies we 
can use the constraint that all Cepheids lie at the same distance,
but this assumption begins to fail for the LMC and may be a poor
assumption for the SMC (e.g. Caldwell \& Coulson 1986, Caldwell \& Laney 1991).  
We allowed the Magellanic Clouds to be tilted
relative to the line of sight, and to have a finite thickness.  
Distances and extinctions were fit for each Galactic Cepheid, constrained by
the non-linear likelihood of fitting the measured radial velocity with a flat 
rotation curve.  From the rotation curve model we obtain estimates of the
solar radius $R_0$ and circular velocity $\Theta_0$.  
We model the priors using Gaussians 
with mean values of $\langle \mu \rangle$ for the distance modulus, 
$\langle E \rangle$ for the extinction, and $\langle Z \rangle$ for the
metallicity, with dispersions of $\sigma_\mu$, $\sigma_E$, and $\sigma_Z$
for each galaxy or Cepheid.  The relative calibration of the standard 
and the HST V and I magnitudes is uncertain at the level of 0.05 mag (Hughes et al. 1994), so we include 
two HST calibration variables constrained by a Gaussian prior of width 0.05 magnitudes
in the likelihood.  The temperature $\sigma_{PC}$ and
PLC $\sigma_{PLCk}$ widths were the same for all galaxies.  

We can divide our analysis into two parts.  First, we
estimate the parameters of the individual Cepheids   
($\mu$, $\delta t$, $E$, and $Z$) given the current parameters of the global
model.  Second, we optimize the parameters ($V_{0k}$, $\alpha_k'$, $\beta_k$,
$\gamma_k'$, $\cdots$) of the global model. 
For a particular Cepheid we have mean magnitude measurements $V_{mk}$ with 
uncertainties $\sigma_{mk}$ in each of $k=1 \cdots N$ bands and a 
known period $P$.  If we define $\sigma_k^2 = \sigma_{mk}^2 + \sigma_{PLCk}^2$, then the
log-likelihood for the model to fit the measured mean magnitudes $V_{mk}$ of 
a particular Cepheid is
\begin{equation}
   -2\ln L = \sum_{k=1}^N { (V_{mk} - V_k)^2 \over \sigma_k^2 } +
  { \delta t^2 \over \sigma_{PC}^2} 
  + { (E - \langle E \rangle)^2 \over \sigma_E^2 }
  + { (\mu - \langle \mu \rangle)^2 \over \sigma_\mu^2 }
  + { (Z - \langle Z \rangle)^2 \over \sigma_Z^2 } + \ln |S^{-1}| 
\end{equation}
up to a constant, where the covariance matrix is
\begin{equation}
  S^{-1}_{ij} = \sigma_i^2 \delta_{ij} +  \sigma_\mu^2 d_i d_j + 
     \beta_i\beta_j \sigma_{PC}^2 + R_i R_j \sigma_E^2 +
     \gamma_i' \gamma_j' \sigma_Z^2,
\end{equation}
and the indices run over the filters included in the calculation.
We must use the determinant of the covariance matrix $S$ in the likelihood
if we are to simultaneously determine the properties of the individual
Cepheids and their global statistical relations.

We can relate our model to standard analyses by breaking the calculation into two
sections: first, deriving the deviations in the properties of individual Cepheids from
the mean, and second, determining the PLC relation vectors and the mean properties
of the Cepheids.  If we measure the magnitude residuals 
$\Delta V_k = V_{mk}-\langle V_k\rangle$ relative to the mean PL relations (eqn. (3)) 
for the parent galaxy, define the deviation of the Cepheid parameters from the mean 
parameters by $x$ and the vector-weighted residuals by $v$,
\begin{equation}
   x = \left(
       \begin{array}{c}
           \delta t  \\
           E - \langle E \rangle \\
           \mu - \langle \mu \rangle \\
           Z - \langle Z \rangle 
         \end{array}
         \right)
   \qquad\hbox{and}\qquad
   v = \left(
       \begin{array}{c}
           \sum_k  \Delta V_k \beta_k   / \sigma_k^2   \\
           \sum_k  \Delta V_k  R_k      / \sigma_k^2   \\
           \sum_k  \Delta V_k  d_k      / \sigma_k^2   \\
           \sum_k  \Delta V_k \gamma_k' / \sigma_k^2 
         \end{array}
         \right),
\end{equation}
and define the matrix $C$ by
\begin{equation}
   C = \left(
       \begin{array}{cccc}
            \sum_k { \beta_k^2 \over \sigma_k^2 } + { 1 \over \sigma_{PC}^2 }
           &\sum_k { \beta_k R_k       \over \sigma_k^2 }
           &\sum_k { \beta_k d_k       \over \sigma_k^2 }
           &\sum_k { \beta_k \gamma_k' \over \sigma_k^2 }  \\
            \sum_k { \beta_k R_k       \over \sigma_k^2 }
           &\sum_k { R_k^2 \over \sigma_k^2 } + { 1 \over \sigma_E^2 }
           &\sum_k { R_k d_k           \over \sigma_k^2 }
           &\sum_k { R_k \gamma_k'     \over \sigma_k^2 }  \\
            \sum_k { \beta_k           \over \sigma_k^2 }
           &\sum_k { R_k d_k           \over \sigma_k^2 }
           &\sum_k { d_k^2 \over \sigma_k^2 } + { 1 \over \sigma_\mu^2 }
           &\sum_k { d_k  \gamma_k'     \over \sigma_k^2 }  \\
            \sum_k { \beta_k \gamma_k' \over \sigma_k^2 }
           &\sum_k { R_k \gamma_k'     \over \sigma_k^2 }
           &\sum_k { d_k \gamma_k'     \over \sigma_k^2 }  
           &\sum_k { \gamma_k'^2       \over \sigma_k^2 } + { 1 \over \sigma_Z^2 }
         \end{array}
         \right)
\end{equation}
then $ x = C^{-1} v$ and the covariance matrix of the parameter estimates for 
a fixed PLC relation is $C^{-1}$.   With the optimization of these variables,
the contribution of the Cepheid to the likelihood becomes
\begin{equation}
    L \propto |S|^{-1/2} 
    \exp( - { 1 \over 2 } \Delta V^T S \Delta V )
\end{equation}
where $\Delta V$ is the vector of residuals relative to the mean
PL relations and $S$ is the covariance matrix defined in eqn. (5).

We obtain the standard extragalactic method (see Madore \& Freedman 1991) if
the four priors have zero width ($\sigma_{PC}=\sigma_E=\sigma_\mu=\sigma_Z=0$) 
and there is no metallicity dependence ($\gamma'_k=0$).  The covariance
matrix $S_{ij} = \delta_{ij} (\sigma_{mi}^2+\sigma_{PLCi}^2)$ is diagonal.
Some extragalactic Cepheid distances are derived by using reddening free
distances or determining individual extinctions (e.g. Freedman et al. 1990, 1991, 1992,
Tanvir et al. 1995, Saha et al. 1996ab, 1997), which corresponds
to allowing $\sigma_E \rightarrow \infty$.
If we optimize the widths of the priors, then the method matches that of Gould (1994) 
or Sasselov et al. (1996) in using the covariance matrix of the residuals to build a better 
statistical model of the data.  The advantage of our formalism is that it derives
from a physical model and we gain new physical insights into the 
system from the variables making up the covariance matrix.  The disadvantage
is that if our physical model is incorrect or it is missing important 
sources of correlated variance in the data, it is not as optimal or
correct a statistical approach as using a purely empirical covariance
matrix. If we limit the model to two or three filters and use broad priors
we recreate the normal Galactic approach (e.g. Caldwell \& Coulson 1986, 1987,
Laney \& Stobie 1986, 1993, 1994, Fernie 1990, Fernie et al. 1995) with the
color dependence of the standard PLC relations appearing indirectly 
through the temperature variable.

Three classes of data and parameters enter the problem.
The first class consists of the period and the composition. 
Here we know the dependent variable ($p$ or $Z$), 
and seek to determine the state vectors ($\alpha'_k$ and $\gamma'_k$).  Given
an adequate range for the dependent variables in an external galaxy, the state
vectors are well-determined and non-degenerate. In particular, the standard PL relations
(e.g. Madore \& Freedman 1991, Laney \& Stobie 1994, Tanvir 1996) 
determine $\alpha'_k$ based on the Cepheids in the LMC.
In the second class, consisting of the extinction and the distance,
we know the state vectors ($R_k$ and $d_k$) and would like to determine the
dependent variable ($E$ and $\mu$).  The distance vector is known exactly ($d_k\equiv 1$),
and the extinction vector $R_k$ is known approximately (see \S2.2).  With accurate measurements
in a sufficient number of bands, we can determine the distances and extinctions for
individual Cepheids.  The uncertainties will depend on how well we can determine
the effects of temperature and metallicity, but the formalism will correctly include
these uncertainties in the distance and extinction estimates.  

In the third case, the temperature, we know neither the dependent variable $t$ nor
the state vector $\beta_k$.  One immediate consequence is the existence of a
mathematical degeneracy under a rescaling of the temperature by $t\rightarrow \xi t$.
The likelihood is unchanged if we rescale the other temperature related variables
by $\beta_k \rightarrow \xi^{-1} \beta_k$, $\delta_P \rightarrow \xi \delta_P $, 
$\delta_Z \rightarrow \xi \delta_Z $ and $\sigma_{PC} \rightarrow \xi \sigma_{PC}$.
No observable (i.e. magnitude) depends on the rescaling, and it means that we
cannot set an absolute temperature scale.  A more fundamental problem is that values
found for variables such as the $t$--$\beta_k$ pair may not have physical meaning
assigned to them in the mathematical model.  The minimization procedure will
simply use them to absorb as much variance as possible from the residuals, and
if $\beta_k$ is unconstrained it assumes the value of the most important unmodeled principal 
component of the true covariance matrix.  Only if the primary source of the variance is the 
temperature distribution at fixed period will $t$ represent the physical 
temperature.   In essence, we are defining the covariance matrix in terms
of a few principal components defined by the state vectors, and we assert that the
$\beta_k$ principal component represents temperature.  The degeneracy 
only affects our interpretation of the variables, and the statistical model will still
reproduce the correct observational PLC/PC/PL relations.   For practical purposes
we solved the degeneracy problem by using a prior for the $\beta_k$ derived from fitting Cepheid 
light curves, which we discuss in paper II (Kochanek 1997b).  The difficulty 
in estimating $\beta_k$ (or its equivalent slope in standard PLC relations) and
separating temperature from extinction has lead the extragalactic Cepheid community
to avoid modeling the temperature distribution (e.g. see the critique of PLC relations
in Madore \& Freedman 1991).

\subsection{Data}

We simultaneously analyzed the Cepheids of 17 galaxies (see Table 1), including only 
Cepheids with periods between 7 days and 80 days.  We used all available Johnson UBV, 
Kron-Cousins RI, and Glass-Carter JHK mean magnitudes for the Cepheids, because
the systems with many color photometry (the Galaxy, LMC, SMC, M~31, M~33, and NGC~300) offer the
best hope of separating the effects of distance, temperature, extinction, and metallicity.
With two-color photometry there is no simple means of separating the effects of temperature
and extinction, while with eight-color photometry it should be possible.  
We assigned magnitude uncertainties of 0.05 mag for the Galactic, LMC and SMC Cepheids,
and 0.10 mag for the HST Cepheids and the poorly sampled ground-based data on M~31, M~33, and NGC~300.
Estimates of the distances and extinctions were little affected by changes in the
estimated measurement errors.    
Although we spot-checked many of the inferred magnitudes and
periods for the Cepheids, we used the published periods and magnitudes in our analysis.

\begin{description}
 \item{Galaxy:} We used the Cepheids with radial velocities in Pont et al. (1994) and periods longer
    than 7 days.  Intensity mean magnitudes were checked from the data compilations by Welch (1996) and
    Berdnikov (1987, 1996), and matched existing B and V tabulations (e.g. Fernie et al. 1995).  
    We dropped the Cepheids AA Ser, XZ Car, and SU Cru from the Pont et al. (1994) sample
    as outliers.
 \item{LMC:}  We used the UBVI data from Caldwell (1996), the JHK data from Laney \& Stobie (1986, 1993, 
    1994) which includes earlier data by Welch et al. (1987), and the R data from Madore (1985).  The
    Caldwell (1996) mean magnitudes include earlier data whose sources are reviewed in Madore (1985). 
    We rejected HV2301, HV2378 and HV2749 from our fits as outliers
    in the likelihood. Caldwell \& Laney (1991) and earlier authors also report that these three
    Cepheids have abnormal properties.
 \item{SMC:}  We used the BVRI data from Caldwell (1996) and the JHK data from Laney \& Stobie 
   (1993, 1994).  
    We rejected HV854, HV1369, HV1438, HV1482, HV1484,  and HV1695 as outliers in the likelihood.
    Caldwell \& Laney (1991) and earlier authors report HV1369, HV1484, HV1636, and HV1641 as
    outliers.  We find nothing peculiar about HV1636, and HV1641 was not included in the sample
    from Caldwell (1996).
 \item{M~33:}  We use the BVRI data from Freedman, Wilson \& Madore (1991) excluding 21979, 23764, and
    B1.  The photometry for V12 in Table 5 of Freedman et al. (1991) is shifted to the left by
    one column.  None of the remaining Cepheids stood out in the likelihood distribution. 
 \item{M~31:}  We use the BVRI data from Freedman \& Madore (1990, Freedman 1996).  We rejected
    the 18.5 day period Cepheid in Baade's Field III.
 \item{NGC~300:} We use the BVRI data from Freedman et al. (1992).  We rejected V21 as an
    outlier.  Freedman et al. (1992) comment that the light curve of V21 is not
    well determined.
 \item{M~81:}  We use the VI data from Freedman et al. (1994).
    We left the magnitude calibrations unchanged.  We rejected C8 as an outlier.
 \item{M~100:} We use the VI data from Farrarese et al. (1996).
    There is a typographical error in the distance modulus given in the abstract of 
    Farrarese et al. (1996, erratum 1997), where the correct value is $\mu = 31.04\pm 0.17$. 
    We added the 0.05 mag ``long exposure correction'' to the tabulated Cepheid
    magnitudes (Hughes et al. 1994).  We rejected the two shortest period Cepheids, C68 and C70 as
    outliers.
 \item{M~101:} We use the VI data from Kelson et al. (1996) including all the I photometry 
    (the ``weak photometry restriction'').  The light curve data tables of Kelson et al. 
    (1996) have the time and magnitude columns in different orders.  We rejected C28 
    as an outlier.
 \item{IC~4182:} We use the VI data from Saha et al. (1994).  We rejected C3-V9 as an
    outlier.
 \item{NGC~5253:} We use the VI data from Saha et al. (1995).  We rejected C3-V1 and C3-V2 as 
    outliers.
 \item{NGC~4536:} We use the VI data from Saha et al. (1996a).  Following the authors we included 
    only Cepheids with periods longer than 20 days and quality class 4-6 in the standard 
    analysis.  We added the 0.05 mag ``long exposure correction'' to the tabulated 
    Cepheid magnitudes.  We rejected C3-V12, C3-V16, C3-V24, and C3-V31 as outliers.
 \item{NGC~925:} We use the VI data from Silberman et al. (1996).  We rejected 3--11, 3--29, 4--68,
    and 4--71 as outliers.
 \item{M~96:}  We use the VI data from Tanvir et al. (1995).
 \item{NGC~3351:} We use the VI data from Graham et al. (1996) but restricted our sample to 
    periods longer than 10 days. We rejected C7, C17, C23, and C46 as outliers.
 \item{NGC~4496A:}  We use the VI data from Saha et al. (1996b).  Following the authors we included 
    only Cepheids with periods longer than 17 days and quality class 4-6 in the standard 
    analysis.  We added the 0.05 mag ``long exposure correction'' to the tabulated Cepheid
    magnitudes.
 \item{NGC~4639:}  We use the VI data from Saha et al. (1997).  Following the authors we included 
    only Cepheids with periods between 20 and 63 days and quality class 4-6 in the standard 
    analysis.  We added the 0.05 mag ``long exposure correction'' to the tabulated Cepheid
    magnitudes.
\end{description}

\subsection{Extinction}

We require an extinction vector $R_k=A_k/E(B-V)$ defined for all eight filters.  Our standard
vector is $R_{0k} = (5.05, 4.31, 3.30, 2.73, 2.07, 0.95, 0.64, 0.39)$
for the UBVRIJHK filters based on the Cardelli, Clayton
\& Mathis (1989) model for the extinction curve.  We fixed $R_V=3.3$ to match
the Key Project, but as emphasized by Cardelli et al. (1989) the difference
between $R_V=3.1$ and $R_V=3.3$ is mainly in the absolute normalization of the $R_k$ 
vector and the estimated $E(B-V)$.  The shape of the extinction vector changes very little.
The Type Ia project uses $A_V/A_I=1.7$ instead of $A_V/A_I=1.6$.  
The earlier M~31 (Freedman \& Madore 1990) and M~33 (Freedman et al. 1991) studies used 
$R_V=3.1$ combined with the Cardelli et al. (1989) extinction curve, 
while $R_V=3.3$ was used for NGC~300 (Freedman et al. 1992).  Laney \& Stobie (1993) estimated
that the JHK values were $(0.82, 0.49, 0.30)$.  We do not include the temperature
variations in the extinction coefficient used by many of the Galactic Cepheid models (e.g.
Caldwell \& Coulson 1986, Laney \& Stobie 1986, 1993, 1994, Feast 1987, Fernie 1990).

Cepheid distances are sensitive to the assumed structure of the extinction 
vector $R_k$, but most existing treatments of Cepheid distances treat
the extinction as a known, understood property of the ISM even while
different analyses select different models.  Exceptions are Laney \& Stobie (1993, 1994)
and Sasselov et al. (1996), who systematically varied the $R_k$ or tried to determine
them from the data, and Freedman et al. (1990, 1991) who examined the effects of using
standard $R_V=3.1$ or $3.3$ or extremal models on the distance estimates.  We will use
a fixed extinction model for all the Cepheids, but we allow the
coefficients to adjust themselves to best fit the data, and the uncertainties
in the extinction vector will be included in the distance uncertainties.
We constrain the extinction vector $R_k$ to fit the Cardelli et al. (1989) 
extinction vector $R_{0k}$ by adding a Gaussian 
prior to the likelihood $\propto \exp(-(R_{Ok}-R_k)^2/2\sigma_R^2)$ with $\sigma_R=0.1$
and $R_{0V}=R_V$.  The uncertainty roughly matches the uncertainties in the Cardelli et al.
(1989) model and the differences between various extinction models used
for Cepheid distances.  

The extinction is not a free variable because it must be positive definite
and larger than the estimated foreground extinction (taken from Burstein \& Heiles (1984), see
Table 1).  Freedman et al. (1990, 1992) were the first to face this problem when they
found a negative estimated extinction for the Cepheids in M~31 (Baade IV field) and 
NGC~300, although Sasselov
et al. (1996)  were the first to use the positivity of the extinction as a constraint when
estimating the metallicity dependence of the Cepheid distances from the EROS sample
(Beaulieu et al. 1995) of LMC and SMC Cepheids.  Most treatments have either 
advocated raising the mean LMC extinction (Freedman et al. 1992, B\"ohm-Vitense 1997),
ignored the problem (e.g. Saha et al. 1997), or used a statistically
incorrect solution in setting $E=0$ without changing the distance.  When we search for solutions 
with physical extinctions, we do so by adding a term to the logarithm of the likelihood (eqn. (4)) of 
the form $(E_{ki}-E_{fk})^2/\sigma_E^2$ if the extinction $E_{ki}$ of Cepheid $i$ in galaxy $k$
is less than the estimated foreground $E_{fk}$, with the scale of $\sigma_E=0.045$
set to include the uncertainty
in the foreground extinction and the mean LMC extinction.  Our choice for the penalty
function has the advantage of simplicity, but it rises too abruptly to be an ideal model
for global uncertainties in the extinction scale.  The structure of the penalty
determines the weight attached to Cepheids whose magnitude uncertainties produce 
negative extinction estimates even if the true extinction is positive.
We experimented with several more complicated and mathematically graceful forms for the 
penalty, but changes in implementation had little effect on the physical results.  

\subsection{Metallicity}

There is no debate about the qualitative effects of metallicity on the Cepheids: high
metallicity Cepheids are brighter and cooler than low metallicity Cepheids at fixed
period.  The
debate centers only on the magnitude of the effect on distance estimates.
The flux is reduced in the blue due to line blanketing, and increased in the red 
and infrared by backwarming.  Theoretical estimates suggest the effect is moderate, 
about $(\gamma'_B+\gamma'_V)/2 \simeq 0.14$ mag/dex and $(\gamma'_B-\gamma'_V)=0.16$ mag/dex
for the mean B and V band luminosity and color  (Stothers 1988).
Chiosi et al. (1993) find a weaker effect of about $(\gamma'_V+\gamma'_I)/2=-0.05$ 
to $0.13$ mag/dex in the mean magnitude, and $(\gamma'_V-\gamma'_I)=0.05$ 
to $0.09$ mag/dex in the $V-I$ color.  An obvious signature of a metallicity 
dependence in the observational data should be a correlation of Cepheid
extinction estimates with the metallicity of the parent galaxy. 

Recent attempts to measure the metallicity effects either examined M~31 or 
compared the LMC and SMC.  Freedman \& Madore (1990) examined the Cepheids in 
M~31 where the work of Blair, Kirshner \& Chevalier (1982) 
implied a steep metallicity gradient.  They found a distance change with metallicity of
$\delta \mu \simeq (0.32\pm0.21)$ mag/dex where the numerical estimate
of their gradient is due to Gould (1994).  The weak significance of the 
Freedman \& Madore (1990) estimate is used by the Key Project as the basis
for neglecting the effects of metallicity in distance estimates pending an
analysis of metal rich and metal poor Cepheids in M~101.  Gould (1994) also reanalyzed 
the M~31 data including the empirical covariance matrix of the residuals to find a larger
correction of $(0.88\pm0.16)$ mag/dex, although the numerical value depended  
on the colors used.  The low metallicity Baade IV field required a large negative 
extinction, suggesting that the metallicity requires a color term, as emphasized by
Stift (1995) and Sasselov et al. (1996), unless the mean LMC extinction is significantly
underestimated (Freedman et al. 1992, B\"ohm-Vitense 1997). 
Sasselov et al. (1996) analyzed the EROS sample of fundamental and overtone Cepheids 
in the LMC and SMC (Beaulieu et al. 1995) using a method based on Gould (1994)
but including both distance and color terms for the metallicity.  Unlike the M~31 studies,
where the estimate depends on Cepheids of differing metallicity at a common distance, 
the Sasselov et al. (1996) value relies on positivity of
extinction and estimates of the foreground and internal extinctions of the LMC and
SMC to estimate the effect.  They find a correction of $0.4_{-0.2}^{+0.1}$ mag/dex in
the mean magnitude and $(0.20\pm0.02)$ mag/dex in the V--I color.
The color dependence is similar
to the Caldwell \& Coulson (1986, 1987) and Gieren et al. (1993) estimates for the
Galactic Cepheids, of $(B-V) \propto (0.29\pm0.05)$ mag/dex and
$(V-I)\propto (0.20\pm0.05)$ mag/dex.

We use the logarithmic abundance of oxygen relative to the LMC (see Table 1, 
$\Delta Z=[O/H]-[O/H]_{LMC}$) as 
our metallicity variable because it is the only measured abundance for most of the Cepheid galaxies. 
Where possible we include the spatial gradients of the metallicity from Zaritsky, Kennicutt \& Huchra
(1994) and assign metallicities to the Cepheids based on their positions in the galaxy (M~33, M~31,
NGC~300, M~81, M~101, NGC~925, NGC~3351, M~100).  Where no gradient was available we used a mean metallicity 
for the whole system (LMC, SMC, IC~4182, NGC~5253, M~96, NGC~4536, NGC~4496A).  
NGC~4536, NGC~4496A, and NGC~4695).  We assigned the Galactic Cepheids a metallicity $0.3$ 
higher than that of the LMC, and we decided not to force a radial
metallicity gradient in our current model. For consistency with Freedman \& Madore (1990) and Gould (1994) we used the 
Blair et al. (1982) gradient for M~31 rather than the Zaritsky et al. (1994) gradient.  
Zaritsky et al. (1994) averaged the Blair et al. (1982) and Dennefeld \& Knuth (1981) data
even though Blair et al. (1982) strongly disagreed with Dennefeld \& Knuth's (1981) results.
We also find an anomalously high metallicity estimate for the M~33 Cepheids compared to M~31.
Both of these potential problems could lead to an underestimate of the strength
of any metallicity effect.

\section{Statistical Models of Cepheid Magnitudes and Distances}

In this section we build the full model for the Cepheid mean magnitudes, starting
from the standard extragalactic method in Model 0. Since the residuals are roughly
parallel to the extinction vector, Model 1 allows all Cepheids
to have individually estimated extinctions.  In Model 2 we allow the 
Cepheids to have a distribution in temperature at fixed period, and enforce the positivity of the
extinction on the models.  Finally, we estimate the effects of metallicity in Model 3.  
We used a flat rotation curve model for the Galaxy to obtain 
estimates for the solar radius $R_0$ and circular velocity $\Theta_0$.    
The LMC and SMC are tilted relative to the line-of-sight, have zero
width, and their relative distance is determined between the (bar) centers. 
The tilt is optimized as part of the solution and agrees well with
Caldwell \& Laney (1991).    To facilitate comparisons with the 
standard PL relations of Madore \& Freedman (1991), we assume an LMC
distance modulus of 18.5 mag and a mean LMC extinction of $E(B-V)=0.10$,
but we have shifted the period origin to $\log P_0=1.4$.  We only discuss
the results for the Johnson UBV, Kron-Cousins RI, and Glass-Carter JHK bands.

\subsection{Model 0:  The Standard Method}

We start our analysis by considering the Cepheid correlation functions
in the absence of metallicity ($\gamma'_k=0$), scatter in the temperature
($\beta_k=0$), and allow only the Galactic Cepheids to have individually
determined extinctions.  Model 0 treats all the extragalactic Cepheids using
the standard extragalactic method.  The parameters for the PLC relations and 
their uncertainties are summarized in Table 2, and the derived distances
and mean extinctions for the extragalactic samples are summarized in
Table 3.   Our zero-points and period slopes are in general agreement
with Freedman \& Madore (1991).  The zero-points are approximately 0.04 mag 
and 0.02 mag fainter in the critical V and I bands, but the differences lie well within the
standard uncertainties.  Our PL relation slopes $\alpha'_k$ are uniformly
shallower by $0.05$--$0.10$, although the agreement is again consistent with the
uncertainties in the individual estimates.   The zero-points and 
slopes agree with recent studies by 
Laney \& Stobie (1994)\footnote{
  Laney \& Stobie (1994) found zero-points of $V_{0k}=$ $13.28\pm0.09$,
  $11.90\pm0.06$, $11.47\pm0.06$, and $11.38\pm0.06$ mag and PL slopes
  of $\alpha'_k = $ $-2.87\pm0.07$, $-3.31\pm0.05$, $-3.42\pm0.05$, and
  $-3.44\pm0.05$ in the V, J, H, and K bands for a mean LMC modulus of 18.5 mag}
and Tanvir (1996).\footnote{ 
  Tanvir (1996) found zero-points of $V_{0k}=13.24\pm0.04$ and
  $12.45\pm0.03$ and PL slopes of $\alpha'_k=-2.774\pm0.083$ and
  $-3.039\pm0.059$ for V and I.}
The best fit extinction vector differs from the
nominal Cardelli et al. (1989) vector by $0.15\pm0.08$, $-0.01\pm0.06$,
$-0.07\pm0.05$, $-0.05\pm0.05$, $+0.10\pm0.06$, $+0.02\pm0.06$, and
$0.08\pm0.06$ for U, B, R, I, J, H, and K respectively.  The changes in the 
coefficients are not a simple change in the $R_V$ value of the Cardelli et al. 
(1989) model, and the final uncertainties in the $R_k$ are less than the 
uncertainties in the prior.  The
ratio $A_V/A_I = 1.63\pm0.05$ lies between the Key project ($A_V/A_I=1.6$)
and SN Ia ($A_V/A_I=1.7$) values.  The shifts in the Glass-Carter JHK
coefficients were expected, because we lacked the true mean effective wavelengths
for these filters and had simply set their values to those for the CTIO JHK filters.   
However, the sign of the shift is opposite to that found by Laney \& Stobie (1993),
who found smaller extinction coefficients using a different analysis technique.

Our distance estimates generally agree with the published values (see Tables 3 and 4, 
and Figures 1 and 2a) with a few exceptions.  We find that M~31, M~33, and NGC~300 are 
closer than the published values by $-0.11$, $-0.12$, and $-0.05$ magnitudes respectively.  In M~33 the
reason is the significantly higher extinction estimate of $E(B-V)=0.16$ instead
of $0.10$.  Both earlier Cepheid distance estimates for M~33 (Freedman 1985, Madore et al. 1985), 
and other models tested in Freedman et al. (1991) match our higher 
extinction estimates.  Since the Cepheid distances to M~31 and M~33 calibrate
many other distance indicators,
these shifts would increase some estimates of the Hubble constant by 6--7\%.
Our agreement with the Key Project estimates is excellent, with a mean shift of
$-0.05$ mag, similar to that found by Tanvir (1996) due to zero-point recalibration.
Our error and extinction estimates
are also in good agreement, except for M~81 where we find a significantly higher mean extinction
of $0.075$ (the Key Project value of $0.03$ is less than the estimated foreground
extinction of $0.04$).  For M~96, we agree with the values for the distance and mean
extinction found by Tanvir et al. (1995), but we derive much larger uncertainties of 0.28
(versus $0.16$) mag for the distance and $0.12$ (versus $0.03$) for the extinction.
We found significant disagreements with the Type Ia project in the distance estimates, 
extinction estimates and uncertainties.  On average our distances were $-0.09$
mag smaller, the extinctions were $0.04$ larger, and our distance uncertainties
were generally twice as large.  The results significantly change the zero-point of the MLCS 
SNIa distance scale (Riess et al. 1996), since the distances to the calibrators 1972E in
NGC~5253, 1981B in NGC~4536, and 1990N in NGC~4639 are revised from $28.08\pm0.10$ mag
to $27.70\pm0.32$, $31.10\pm0.13$ to $30.97\pm0.22$, and  $32.03\pm0.22$ to
$32.11\pm0.32$ respectively.  The revised MLCS estimate of the Hubble constant is 
$69\pm8 \kms$ Mpc$^{-1}$, an 8\% increase from the original estimate of $64\pm6 \kms $ Mpc$^{-1}$.
Some of the differences may be due to our use of a consistent extinction law for all the galaxies
(the Type Ia project uses the Madore \& Freedman (1991) PL relation based on 
$R_V/R_I=1.6$ but fit the data using $R_V/R_I=1.7$), and our use of statistically consistent
treatments of the covariance between distance and extinction.  For example, in
NGC~5253 Saha et al. (1995) give large uncertainties in the extinction (see Table 3)
that are statistically inconsistent with the small uncertainties in the distance.  
Figure 2 compares two absolute distance estimates, the expanding
photosphere method for Type II supernovae (Eastman et al. 1996)
and physical models of Type Ia supernovae (H\"oflich \& Khokhlov 1996),
and two relative distance estimates, surface brightness fluctuations
(Tonry 1996) and the MLCS/SNIa (Riess et al. 1996) method, to the
Cepheid distances (see Table 4).

Magnitude selection biases may be important in Model 0 because of the large scatter 
in the PL relations.  Biases have been extensively discussed in the literature, most 
recently by Tanvir (1996), along with debates on how to optimally perform model 
fits (e.g. fitting inverse PL relations as in Kelson et al. (1996)).  We can
avoid the biases almost completely by using PLC relations to model the correlations
in the residuals and to reduce the intrinsic scatter.  Since these approaches
are also physically more useful, we will not discuss selection biases further.
 
\begin{figure}
\centerline{\psfig{figure=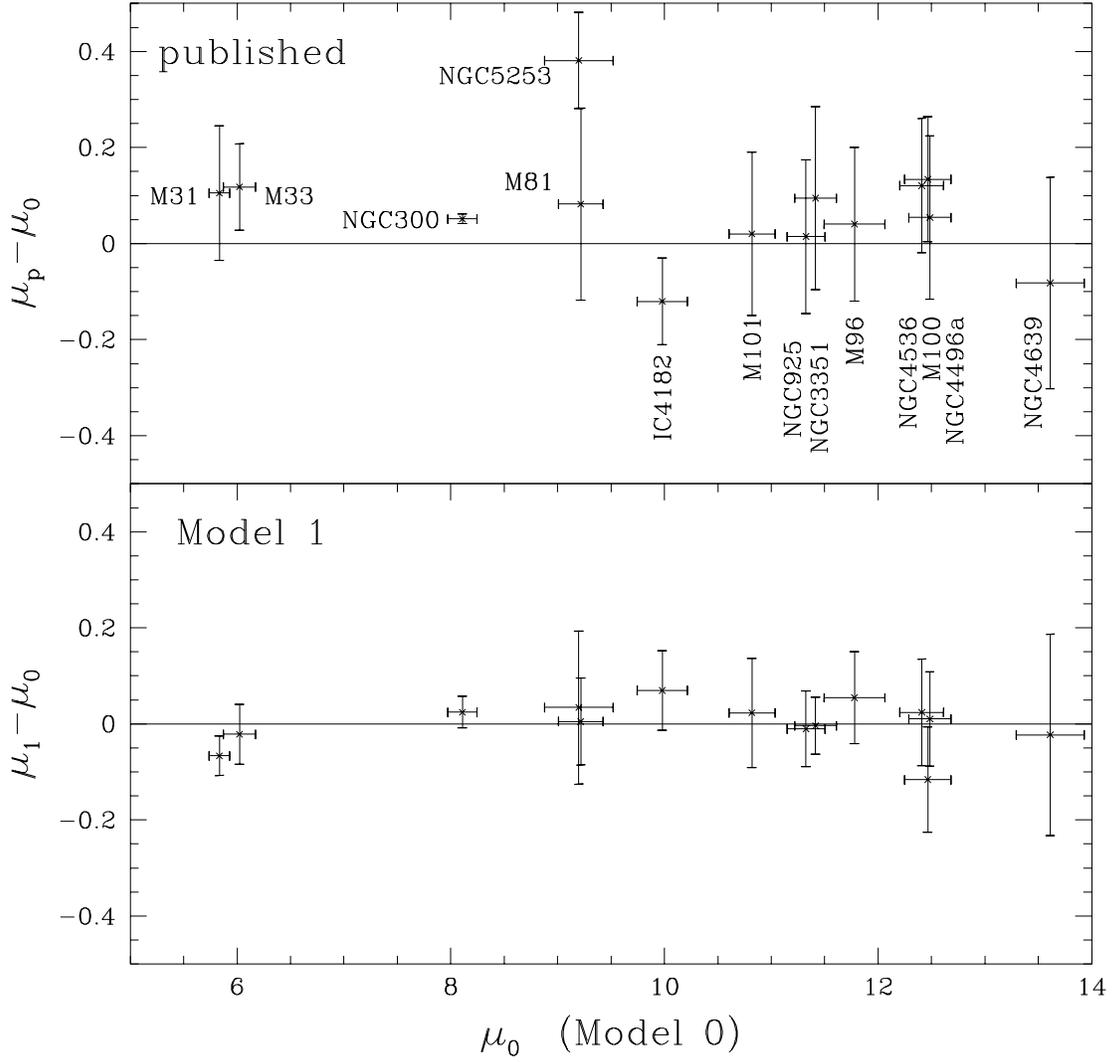,height=6.0in}}
\caption{Distance comparisons between Model 0 and either published (top) or Model 1 (bottom)
  distances.  The horizontal error bar is the error in Model 0, while the vertical error
  bar is the error in the comparison distance. }
\end{figure}

\begin{figure}
\centerline{ \psfig{figure=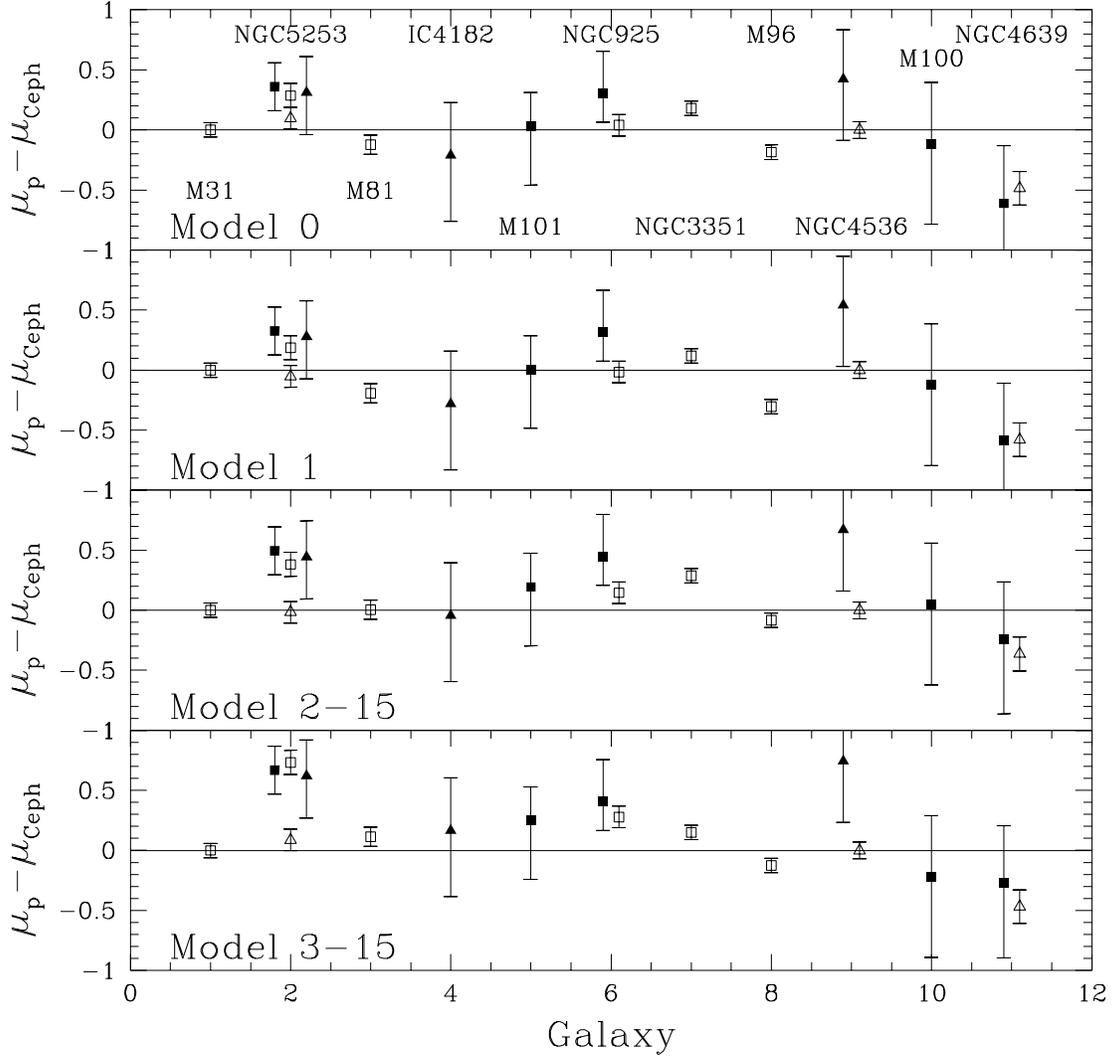,height=6.0in}}
\caption{Distance comparisons to other methods.  The expanding photosphere
  method for SNII (EPM, solid squares, Eastman et al. 1996) and physical models of
  SNIa (SnIa, solid triangles, H\"oflich \& Khokhlov 1996) are absolute distances.
  The surface brightness fluctuation distances are shown relative to M~31
  (SBF, open squares, Tonry et al. 1996) and the SNIa/MLCS method distances
  are shown relative to NGC~4536 (MLCS, open triangles, Riess et al. 1996).  The
  error bars are the uncertainties in the non-Cepheid distance indicator. }
\end{figure}

\begin{figure}
\centerline{\psfig{figure=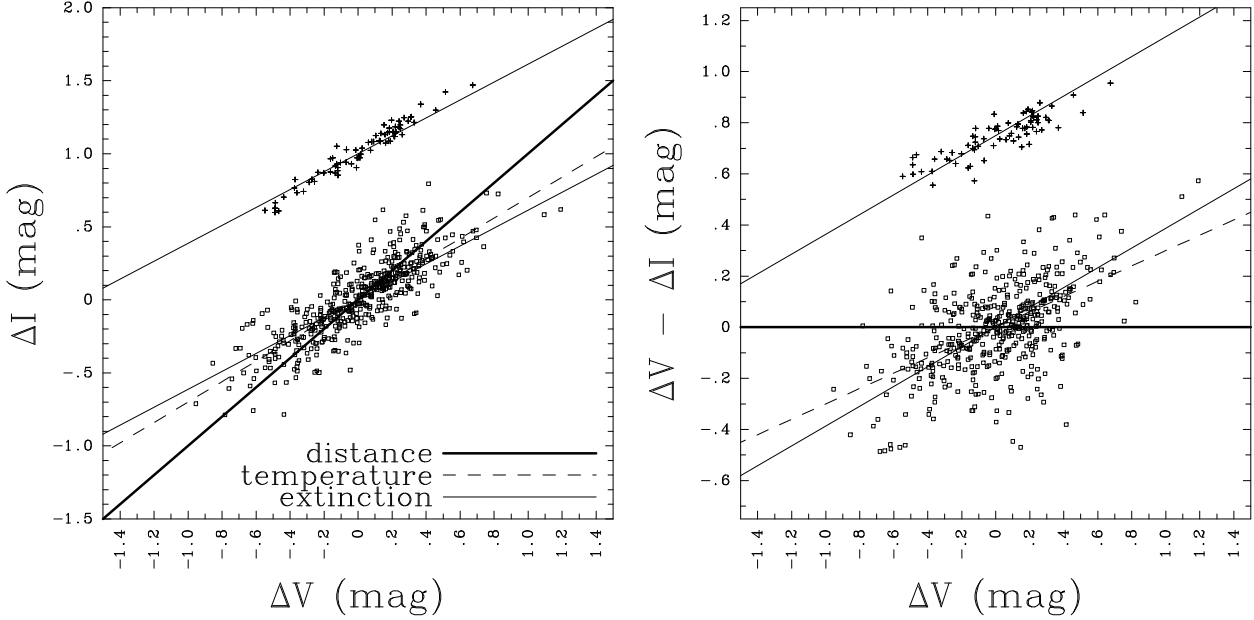,height=3.0in} }
\caption{Magnitude residuals in Model 0.  The left panel is a simple diagram of the $\Delta V$  
  and $\Delta I$ residuals (squares). It shows that the residuals are strongly correlated and lie roughly
  along the extinction vector.   The right panel displays the same residuals using
  the $\Delta V$  versus $\Delta V$--$\Delta I$ format of Type Ia project.  The correlation, while still
  clearly present, is much more difficult to recognize.  We also display the residuals
  (vertically offset, crosses) and the extinction direction for the more accurately measured
  LMC and SMC Cepheids to illustrate the effects of measurement uncertainties on the apparent 
  correlations.  The heavy solid, dashed and light solid lines show the expected directions for 
  distance, temperature and extinction residuals respectively.
        }
\end{figure}

\subsection{Model 1: Scatter In the Extinction}

The magnitude residuals of all the extragalactic Cepheids are highly correlated 
and lie along the extinction direction (see Figure 3).\footnote{Contrary to the conclusions
of the Type Ia project, we find that the residuals in IC~4182, NGC~4536, NGC~4496a and NGC~5253
are also correlated with the extinction vector.  We believe their conclusion 
is an artifact of examining residuals in the space of $\Delta V-\Delta I$ versus $\Delta V$.  
Both the mathematics of the covariance matrix and simple experiments show that when the 
uncertainties in $I$ are larger than those in $V$, the $\Delta V-\Delta I$ versus $\Delta V$
covariance diagrams can look symmetric and uncorrelated even when the data is correlated.  
The quantity $\Delta V - \Delta I$ simultaneously 
reduces the signal from differential extinction and increases the apparent noise (see Fig. 3).}
The importance of allowing individual extinctions is glaringly obvious
if we compare the residuals for the Galactic and extragalactic
samples -- for the Galactic Cepheids there are 624 measured mean magnitudes for 121 Cepheids
with an rms residual of $0.06$ magnitudes, while for the
extragalactic Cepheids there are 1617 magnitudes for 545 Cepheids
with an rms residual of $0.29$ magnitudes.  The independent distance
and extinction estimates for the Galactic Cepheids dramatically
reduce their residuals compared to the rest of the sample even
though a large subset of the Galactic Cepheids have 5 to 8 color
photometry while most of the extragalactic Cepheids have only two
color photometry.  

In Model 1 we allow all Cepheids an independent extinction variable, as
we have already used for the Galactic Cepheids in Model 0.  Each Cepheid 
$i$ in galaxy $k$ was assigned extinction $\langle E_k \rangle + \Delta E_{ki}$, 
where $\langle E_k\rangle$ is the mean extinction and the correction $\Delta E_{ki}$ 
is forced to have zero mean for each galaxy and is limited by a Gaussian prior 
whose width $\sigma_{Ek}$ is simultaneously optimized.  We need the prior because most of the
extragalactic Cepheids have only two filters, and as we
start assigning each Cepheid individual extinctions, temperatures
and metallicities we will overfit the data if we do not include
the covariance matrix (eqns. 4 and 5) and optimize the priors.
Some extragalactic distance estimates have used of individual extinction
estimates (e.g. Tanvir et al. 1995) or reddening free magnitude 
estimates (e.g. Freedman et al. 1990, 1991, 1992, Saha et al. 1996ab, 1997),
although the final distance estimates are usually based on the standard
PL relations.  These individual extinction estimates correspond to taking
the limit $\sigma_E \rightarrow \infty$ in eqns. (4) and (5), and they may
overestimate the width of the extinction distribution in noisy data.

The changes in the Cepheid relations and distances are summarized in Tables 2 and
3 and Figure 1.  In Model 1 the rms residuals for the extragalactic Cepheids drop to 
$0.09$ mag from $0.29$ mag in Model 0.  The rms residuals for the LMC and SMC 
($0.06$ and $0.08$ mag) are smaller than the other extragalactic systems ($0.10$ mag) 
even though the average Cloud Cepheid was measured in 5 filters.
The removal of the extinction-correlated residuals leads to large 
reductions in the PLC relation widths (which represent uncorrelated errors),
but the residuals are still correlated (particularly UJHK) and the PLC widths
are still broader than expected given the estimated measurement errors.
The zero-points and slopes show little change from Model 0 and have
reduced uncertainties, but the estimated extinction vector has changed
considerably in the infrared, suggesting that it has absorbed some of the
variance due to temperature as well as extinction.  In particular, the
distance of the sun from the Galactic center ($R_0$) is strongly affected by the change in the 
IR extinction coefficients (from 7.5 kpc to 6.7 kpc).  

As Figure 3 illustrates, the extinction and temperature are very nearly parallel
for V and I, so it is very difficult to cleanly separate the two physical
terms in noisy two-color data.  Madore \& Freedman (1991) question whether the
two variables can be accurately separated even with more accurate three-color
data used for Galactic Cepheid extinction estimates, although their position
is strongly rejected by Laney \& Stobie (1993, 1994).  Moreover, the HST
Cepheid data may have correlated systematic errors due to crowding, differences
in analysis methods, and the construction of the I mean magnitude using the
V light curves for interpolation that Model 1 interprets as extinction variations.
Such systematic errors have no effect on the distance estimates, since correlated
residuals must be modeled for a correct statistical treatment, but they may
strongly affect our interpretation of the scatter in terms of extinction.  Note,
however, that the spread in color in the extragalactic systems is quite comparable
to the spread in the Magellanic Clouds (Figure 3), and it would be truly 
astonishing if the extinction from a large sample of objects randomly chosen
from spiral galaxies failed to show significant scatter.

Other than the value of $R_0$, the distances are little changed from Model 0 (see Fig. 1),
with a mean shift of less than 0.01 mag.  The typical distance and mean extinction uncertainties, 
however, are roughly half those of Model 0,  although the reduction is most dramatic for galaxies 
with small numbers of Cepheids (e.g. M~96).  The change is largely due to the reduced rms magnitude
residual in estimating the intrinsic Cepheid luminosities.  In Model 0
the statistical uncertainty is $\sim 0.29/N^{1/2}$ magnitudes where the 
numerical coefficient is the combination of the measurement errors and the PLC relation width,
while in Model 1 the typical uncertainty is $\sim 0.10/N^{1/2}$ magnitudes.  
The total uncertainties are larger due to 
calibration uncertainties in the PLC relations and the HST magnitudes that
are unaffected by statistical averaging.  Fitting the correlated residuals
also makes our model significantly less sensitive to magnitude selection 
biases than Model 0, although the general agreement of the results suggest they were not of 
great importance.   The MLCS estimate of the Hubble constant becomes $72\pm6 \kms$ Mpc$^{-1}$, slightly
higher than in Model 0.
Figure 2 compares the Cepheid distances to the distances in Table 4.

\subsection{Model 2: Temperature and the Positivity of the Extinction}

Figure 4 shows the mean extinctions relative to the foreground extinction and the width
of the extinction distribution as a function of the metallicity of the host galaxy.
The Model 1 solutions are unphysical because many of the galaxies have either negative
mean internal extinctions (SMC, NGC~300, IC~4182 and NGC~4639) or extinction distributions 
extending to negative internal extinctions.  Even for the LMC and the SMC,
where we have many filters, good accuracy, and no systematic problems
such as the HST calibration terms, there are Cepheids with extinctions less than 
the expected foreground extinction.  The simplest solution to the problem, and the
one advocated by Freedman et al. (1992) when they first faced the problem for the
NGC~300 Cepheids and more recently advocated by B\"ohm-Vitense (1997), 
is simply to raise the mean LMC extinction from the standard $\langle E \rangle_{LMC}=0.10$
assumed in the extragalactic Cepheid analyses.  Both Freedman et al. (1992) and
B\"ohm-Vitense (1997) suggest that the problem can be solved by increasing the 
mean extinction to $\langle E \rangle_{LMC}\simeq 0.18$.  Our larger sample and
the interpretation of the scatter in color as an extinction distribution increases
the required LMC extinction to $\langle E \rangle_{LMC}\gtorder0.25$.  
The standard Galactic Cepheid analyses derive mean LMC extinctions of only $0.07$
(e.g. Caldwell \& Coulson 1986), even lower than $\langle E \rangle_{LMC}=0.10$, 
based on a combination of the Cepheid colors and other estimates of the space reddening.  
Grieve \& Madore (1986) found median LMC supergiant extinctions of approximately 
$0.10$, with 90\% of the estimated extinctions below $E(B-V)=0.18$.
Bessel (1991), in arguing for higher than generally accepted extinctions for the LMC and SMC, 
advocates an increase to only $\langle E\rangle \sim 0.13$ based on HI column
densities, optical polarization, interstellar absorption lines, and color excesses.
Thus the trivial solution to the negative extinction problem of simply raising 
$\langle E \rangle_{LMC}$ can only be a partial solution.   As Freedman et al. 
(1992) also noted, the temperature distribution at fixed period contributes
to the negative extinction problem if the two variables are not distinguishable
and hotter than average Cepheids are assigned reduced extinctions.  

In Model 2 we try to minimize the effects of the temperature distribution in
biasing the extinctions by giving each Cepheid a temperature deviating by $\delta t$ from 
the mean for its period, constrained by a Gaussian prior of width $\sigma_{PC}$ 
corresponding to the width of the instability strip (see eqn. 2).
We also force the extinctions to be positive by
adding the extra terms to the likelihood function discussed in \S2.2.  Since there
is some uncertainty in the extinction normalization, we explore three models with
$\langle E \rangle_{LMC}=0.10$, $0.15$, and $0.20$ labeled by Model 2--10, 2--15,
and 2--20, that correspond to low, slightly high, and very high estimates for the
mean extinction. We adopt Model 2--15 as our standard, and the resulting PLC vectors, 
distances, and extinctions are presented in Tables 2 and 3.
In Model 2 we must determine the temperature vector $\beta_k$ as well as the
temperature corrections, $\delta t$, and as is frequently 
noted in the extragalactic critiques of the Galactic Cepheid methods (e.g. Madore 
\& Freedman 1991), the temperature state vector $\beta_k$ is hard to determine 
uniquely from the mean magnitudes.  
Indeed, we found that we could not stably estimate $\beta_k$ without including a prior 
based on the variation of color with phase.  
Based on an analysis of Cepheid light curves (Kochanek 1997b), we estimated
a model $\beta_{0k}$ (see Table 2), assumed that the uncertainty
in the coefficients was $0.05$, and added a Gaussian prior for the deviation
of $\beta_k$ from $\beta_{0k}$ to the likelihood.  
We should note, however, that many of the Galactic analyses
use a very similar approach to calibrating the PLC relations 
(e.g. Laney \& Stobie 1986, 1993, 1994).  The 
need to make such a strong assumption about $\beta_k$ combined with (or due to) the
limited two-color photometry on most of the extragalactic systems is a 
primary limitation for the remainder of our analysis, but the simultaneous
optimization of the prior widths keeps us from overfitting the data.

\begin{figure}
\centerline{\psfig{figure=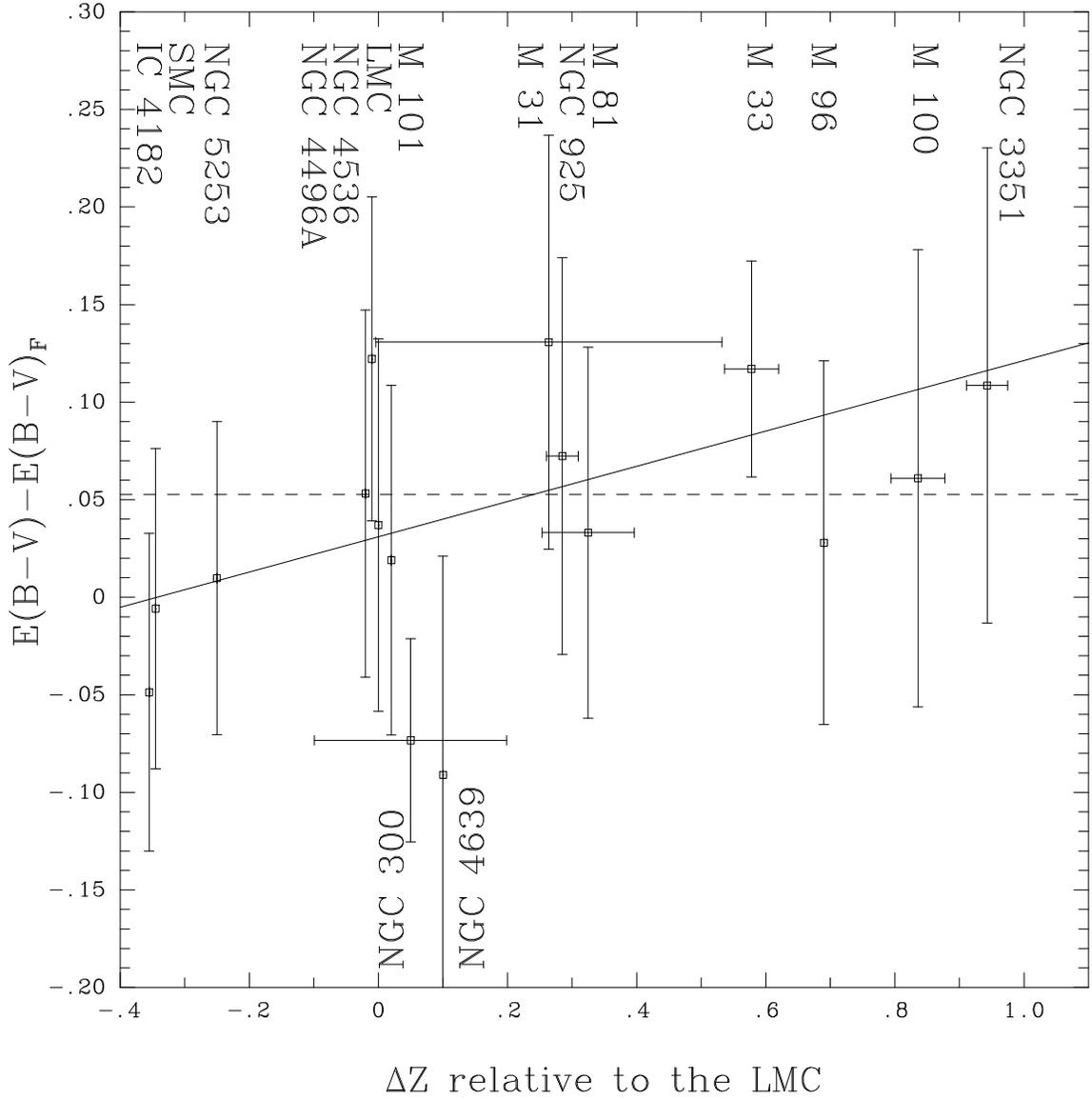,height=6.0in}}
\caption{Extinctions and metallicity.  The points for each galaxy show the mean extinction
  relative to the foreground extinction
  $\langle E_k \rangle-E_{fk}$ and the width of the extinction distribution $\sigma_{Ek}$ for Model 1
  as a function of the metallicity.  The uncertainty in the mean extinction is smaller than
  $\sigma_{Ek}$ by roughly the square root of the number of Cepheids in the galaxy (see Table 3).
  The metallicity error bars show the spread in the
  metallicities of the Cepheids used in the model.  The solid and dashed lines show the
  best linear and constant fits for the trend excluding NGC~300 and NGC~4639.  Several points have been
  shifted slightly in metallicity to separate the vertical error bars.
   }
\end{figure}

\begin{figure}
\centerline{ \psfig{figure=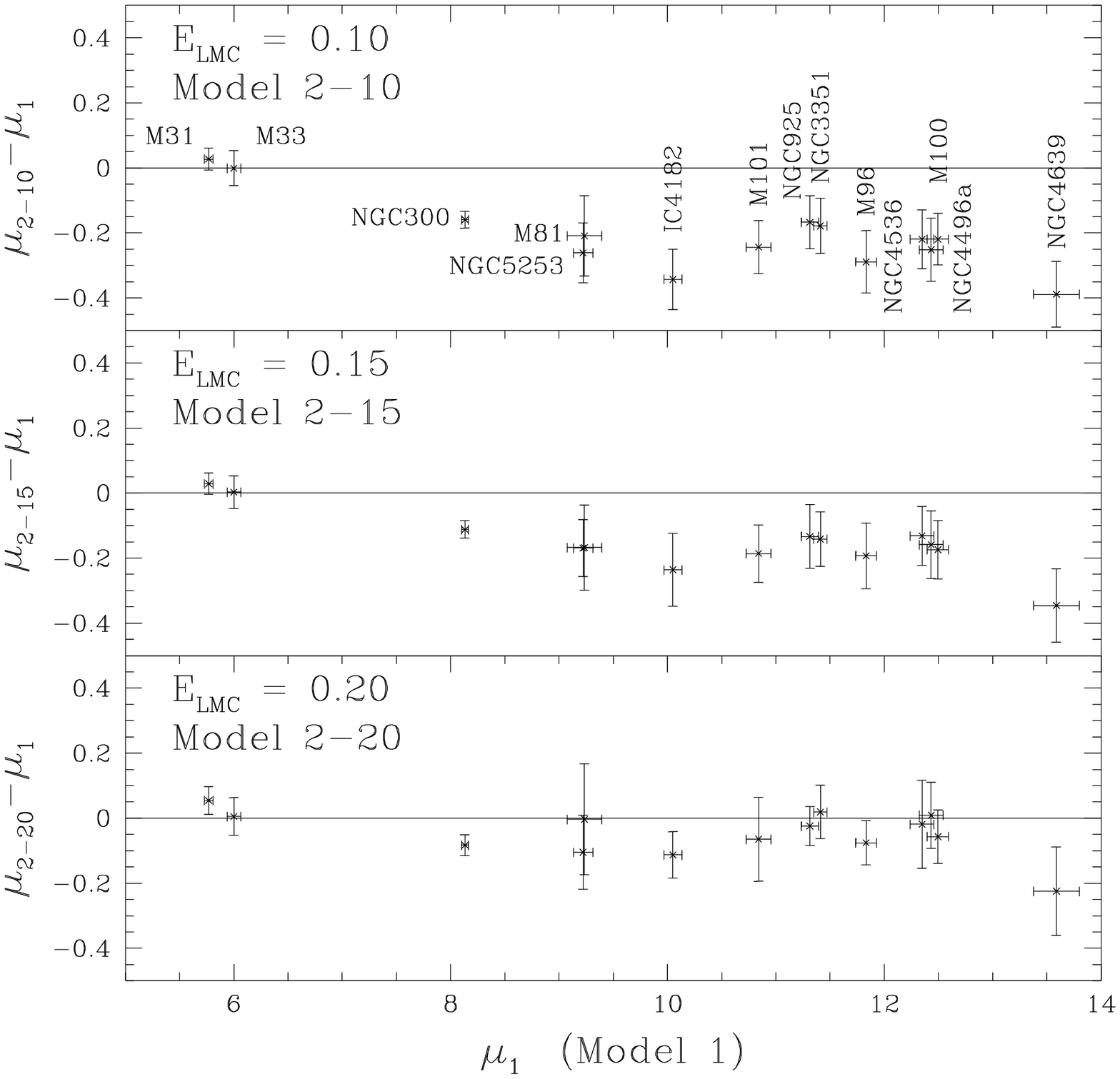,height=6.0in}}
\caption{
Distance comparisons between Model 2 and Model 1 distances for the low (top), middle (center)
  and high (bottom) LMC extinction estimates.  The horizontal error bar is the error in Model 1, 
  while the vertical error bar is the error in Model 2. }
\end{figure}

\begin{figure}
\centerline{\psfig{figure=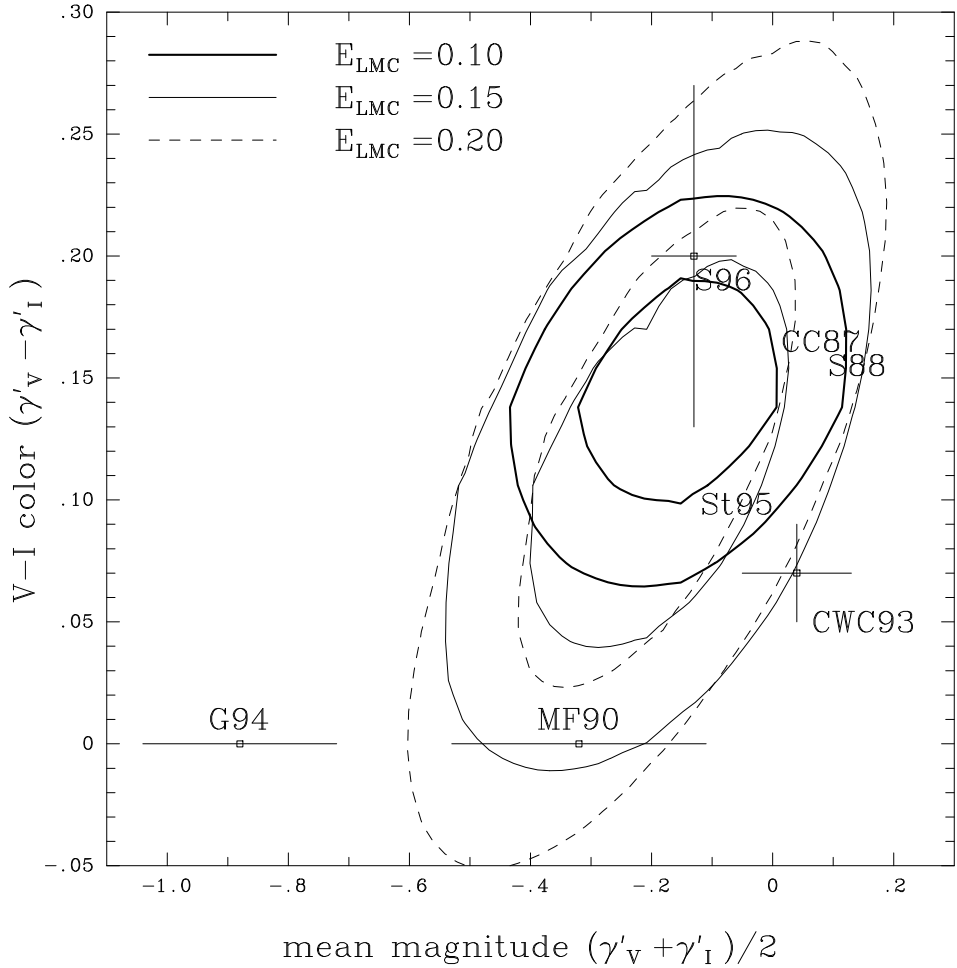,height=5.0in}}
\caption{Metallicity corrections to the zero point.  Likelihood contours for the change in
  mean magnitude ($(\gamma'_V+\gamma'_I)/2$) and V--I color ($\gamma'_V-\gamma'_I$) compared
  to other observational and theoretical determinations.  The contours are the 1--$\sigma$
  ($\Delta\chi^2=2.30$) and 2--$\sigma$ ($\Delta\chi^2=6.17$) confidence intervals for two
  parameters. The G94 (Gould 1994) and MF90 (Madore \& Freedman 1990, as estimated by
  Gould 1994), estimates included no color variation.  CWC93 marks the theoretical estimate
  for V--I from Chiosi et al. (1993).  S88 and St95 mark the theoretical estimates for B--V from
  Stothers (1988) and Stift (1990, 1995).
  CC87 marks the semi-empirical model for B--V from Caldwell \& Coulson (1987).
  S96 marks the Sasselov et al. (1996) V and I determination from the EROS Cepheids. 
   }
\end{figure}

\begin{figure}
\centerline{ \psfig{figure=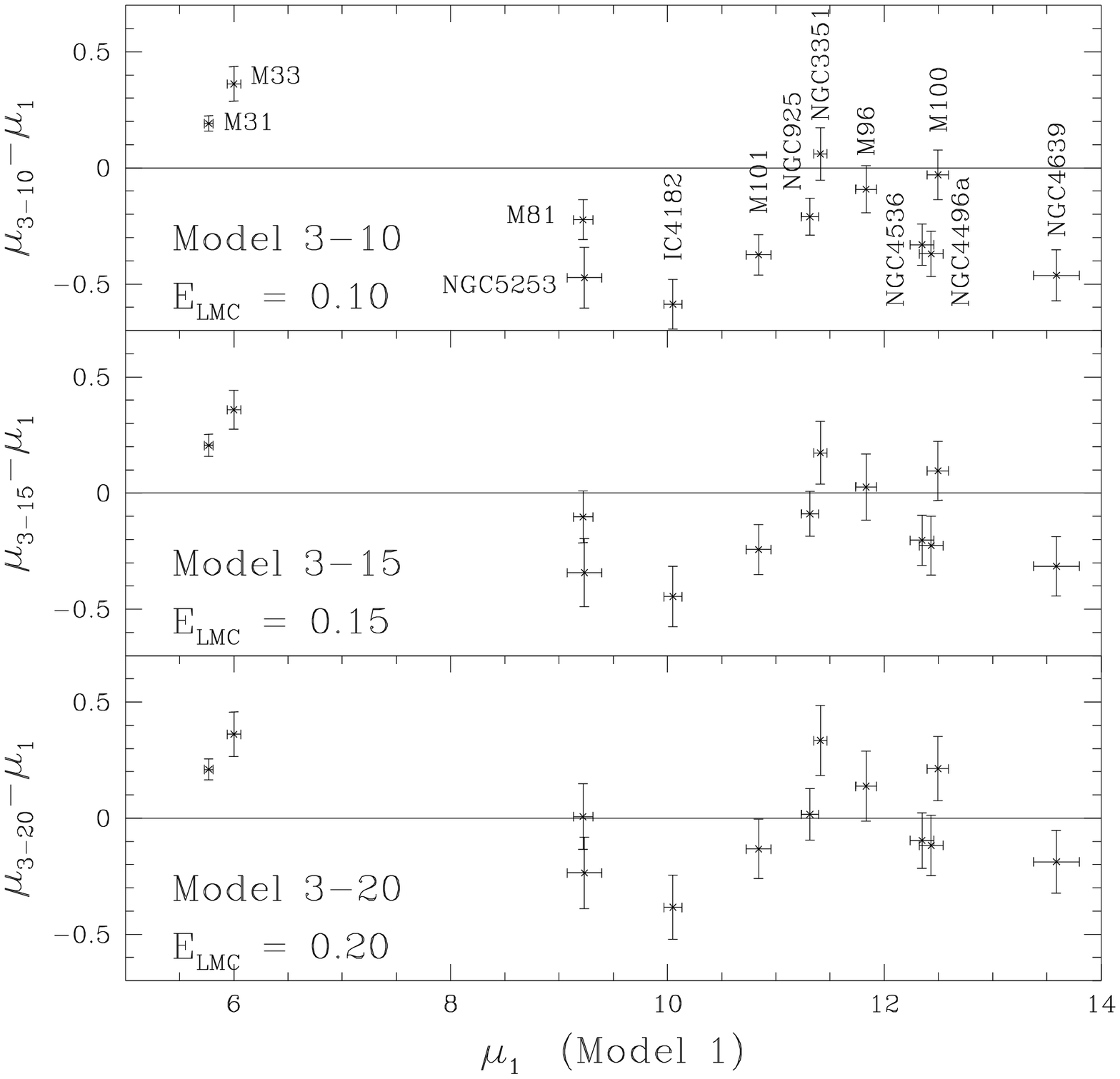,height=6.0in} }
\caption{
Distance comparisons between Model 3 and Model 1 distances for the low (top), middle (center)
  and high (bottom) LMC extinction estimates.  The horizontal error bar is the error in Model 1, 
  while the vertical error bar is the error in Model 3. }
\end{figure}

We find a width for the instability strip of $\sigma_{PC}\simeq0.18\pm0.01$ in
all three models, which corresponds to a FWHM in the V--I (B--I) colors at fixed period 
of 0.08 (0.14) mag.  Unlike the previous two models, Model 2 gives non-zero values 
for HST calibration variables ($\Delta V$,$\Delta I$) of 
($0.05\pm0.04$,$-0.04\pm0.04$), ($0.03\pm0.04$,$-0.03\pm0.04$), ($0.02\pm0.04$,$-0.03\pm0.04$)
for Models 2--10, 2--15, and 2--20 respectively.  In all three cases the calibration
uncertainties provide a significant part of the solution to the negative extinction
problem.  Note that the PLC zero points for Model 2-15 are not directly comparable to the
previous models due to the change in the extinction. The infrared 
extinction coefficients are now closer to the priors than in Model 1 and agree 
with the Laney \& Stobie (1993) estimates and the estimated value of $R_0=7.7\pm0.3$ kpc
is again consistent with other estimates (Reid 1993).

Standard Cepheid distances are {\it biased} distance estimators because they treat positive
and negative extinctions equally.  The sense of the bias is always to overestimate the
distances, because lower extinctions correspond to higher distances.  Random photometry
errors can and will produce negative extinctions for objects with positive extinctions,
but such problems affect only the mathematics of implementing the positive extinction 
condition, not the need for it. 
Any procedure to impose the physical condition that the extinction is positive must reduce
the likelihood for large distances and thereby drive the best fit distance estimate downwards.
It is also a global bias, because the Cepheid distance and extinction estimates are 
tightly correlated by the PLC relations.  When we force a galaxy with strongly negative 
extinction estimates like NGC~300 to have a higher extinction and a lower distance,
the correlations caused by all the Cepheids sharing the same estimate of their intrinsic
colors will also force galaxies like M~33 with strictly positive extinctions to 
higher extinctions and lower distances as well.  {\it Adjusting the extinction of one
galaxy without adjusting the extinction of all galaxies is physically and statistically
incorrect, unless the reason for the adjustment is a systematic error in the data 
for that particular galaxy.}  The negative extinction problems should not be solved ``locally.'' 
    
As Figure 5 shows, Model 2 simply drives the galaxies to lower distances and higher 
extinctions, until it is reasonably unlikely that any Cepheids have negative extinctions.
The mean distances shift by $-0.20$, $-0.14$, and $-0.05$ mag for Models 2--10, 2--15,
and 2--20 respectively, with the low LMC extinction Model 2--10 requiring the largest
shift.  The effect is strongest on the low metallicity galaxies, which means the
changes in the MLCS calibrations are considerable.  The MLCS Hubble constant
estimates are $80\pm6$, $78\pm6$, and $72\pm6\kms$ Mpc$^{-1}$ for the three models
compared to $72\pm6\kms$ Mpc$^{-1}$ for Model 1.  Because the effects of the bias
are global, even a relatively high extinction galaxy like M~100 is driven to 
lower distances. Scaling from the Mould et al. (1995) estimate of $80\kms$ Mpc$^{-1}$, 
which becomes $82\kms$ Mpc$^{-1}$ in Model 1, the Hubble constant shifts to $H_0=90$, 
$88$, and $84\kms$ Mpc$^{-1}$ in Models 2--10, 2--15, and 2--20 respectively.\footnote{The
uncertainties in the $H_0$ estimate from M~100 are dominated by the systematic uncertainties
in the model for Virgo (Mould et al. 1995).}  
Despite the large shifts, the agreement with other distance indicators (see Fig. 2)
is no worse than in Model 1, if only because the absolute distance indicators have
low accuracy while the relative distances are almost unchanged.
The goodnesses of fit for Model 2 are considerably worse than Model 1, even though we
have allowed separate temperature estimates for the Cepheids.  The likelihood function
has increased by $\Delta 2 \ln L = 139$, $78$, and $65$ relative to Model 1, despite
{\it adding} 695 new degrees of freedom to the model!

\subsection{Model 3: Metallicity}

The Model 2 solutions are not very attractive because of the huge downward shifts in the
distances, and the association of unphysical extinctions with low metallicity in Figure 4 strongly
suggests the need for a metallicity dependence in the Cepheid distance.  In Model 3
we add a metallicity dependence to the Cepheid zero-point vector $\gamma'_k$, as in the model
of Sasselov et al. (1996).  The formal trend of extinction with metallicity (neglecting NGC~300 and
NGC~4639) corresponds to a color dependence of $V-I \propto (0.09 \pm 0.01) \Delta [O/H]$, which
is comparable to the color variations predicted in theoretical models (e.g. Stothers 1988,
Stift 1990, Chiosi et al. 1983) and previous experimental estimates (e.g. Gieren et al. (1993),
Sasselov et al. (1996)).  There are, however, two discrepant points in Figure 4, NGC~300
and NGC~4639, both of which are significantly bluer than the trend.  For NGC~4639 the 
large uncertainties in the extinction can probably explain the discrepancies.  NGC~300 is 
peculiar, and we drop it from our subsequent
analysis -- in doing so, we are assuming that the blue color of the Cepheids is either an artifact 
of the Freedman et al. (1992) data or that the HII region abundance estimates are incorrect.

In Model 3 we added the metallicity dependent zero-point correction $\gamma'_k \Delta Z$ to
Model 2, and repeated the calculation for the three different mean LMC extinctions labeled
by Model 3--10, 3--15, and 3--20.  As in Model 2, we present the full results only for 
Model 3--15 in Tables 2 and 3.  We fit the data leaving the input metallicities fixed 
($\sigma_Z \rightarrow 0$), because attempting to determine the metallicities from the
data was unjustified by the amount and quality of the data.  If we have found
a solution for the distance modulus $\mu_1$ and extinction $E_1$ ignoring the
metallicity zero point correction $\gamma'_k$, then the true distance modulus is 
\begin{equation} 
  \mu_0 = \mu_1 - { \gamma'_V + \gamma'_I \over 2 } \Delta Z 
                + { \gamma'_V-\gamma'_I  \over 2 } { R_V +R_I \over R_V- R_I} \Delta Z 
\end{equation}
and the true extinction is
\begin{equation}
  E_0 = E_1 - { \gamma'_V - \gamma'_I \over R_V - R_I } \Delta Z
\end{equation}
where the terms are broken into the change in the mean  V and I magnitudes and the change in the
V--I color.  If low metallicity Cepheids are blue ($\gamma_V-\gamma_I >0$) then 
adding the metallicity term can solve the extinction problem without driving the distances of
all galaxies downwards as in Model 2.  The metallicity terms will not, however, change
the embarrassingly low distances to the low metallicity galaxies in Model 2, because their
distances are still reduced by $(R_V+R_I)/2 \sim 2.7$ times the change in the extinction.  

Our solutions generically made the metal rich Cepheids redder and brighter than the metal 
poor Cepheids, as shown in Figure 6.  The best fit solution changes little with the assumptions
about the LMC extinction, and the absence of a metallicity dependence is ruled out
at greater than 95\% confidence. The mean luminosity changes in the V and I bands,
$(\gamma'_V+\gamma'_I)$, are $-0.13\pm0.09$, $-0.15\pm0.14$, and $-0.17\pm0.15$ mag/dex
for Models 3--10, 3--15, and 3--20 respectively, and the mean changes in color,
$\gamma'_V-\gamma'_I$, are $0.15\pm0.03$, $0.13\pm0.04$, and $0.15\pm0.07$ respectively.
The less strongly we restricted the permitted range for the extinctions, the more uncertain
the metallicity dependence.  As pointed out by Sasselov et al. (1996), the positivity
of the extinction is an important component in the quantitative determination of composition
effects.  The HST calibration uncertainties still represent a major problem, and
they contribute much of the solution to maintaining positive extinctions.  The
calibration variables ($\Delta V$, $\Delta I$) are ($0.09\pm0.04$, $-0.07\pm0.04$),
($0.05\pm0.04$, $-0.05\pm0.04$), and ($0.03\pm0.04$, $-0.03\pm0.04$) for the three
solutions.   The overall structure of the metallicity vector
$\gamma'_k$ matches theoretical expectations.  Metal rich Cepheids show a decreased
flux in U and B, and then a gradually increasing flux in the redder bands, reaching
a plateau in the infrared.  The uncertainties in the $\gamma'_k$ in Table 2
are dominated by the uncertainties of the mean luminosity change, and as the error
ellipses in Figure 6 demonstrate, the uncertainties in the color changes are
considerably smaller. 

Most of the debate about composition effects on extragalactic Cepheid distances 
has focused on the M31 Cepheids studied by Madore \& Freedman (1990) and Gould (1994).  
These models allowed no color changes due to composition
($\gamma'_k=\gamma'_j$), while it is clear both from theoretical models,
previous experimental estimates, and our estimates, that the changes in 
color are the dominant source of changes in distance estimates.  Our estimates 
of the effect are slightly higher than the theoretical estimates of Chiosi et al. (1993).
The B--V color change of $0.28$ mag/dex is not as well constrained because 
most of the galaxies lack B photometry, but it is larger than the
theoretical estimates of Stothers (1988) and Stift (1990, 1995).
The values match the experimental determination by Sasselov et al. (1996)
using the EROS sample of Cepheids in the LMC and SMC and the
typical Galactic Cepheid metallicity correction models (e.g. Caldwell \& Coulson 1986, 1987,
Gieren et al. 1993).

Figure 7 shows the changes in the distances relative to Model 1 for the three different
assumptions about the LMC extinction.  As expected, the metal poor galaxies
(e.g. NGC~5253, IC~4182, NGC~4536) are shifted to lower distances relative
to the metal rich galaxies (e.g. NGC~3351, M~33, M~100). The mean change
in distance depends on the mean extinction of the LMC, with Model 3--10
still requiring a significant reduction in the mean distances.  The MLCS
estimates of the Hubble constant become $85\pm6$, $80\pm6$ and $79\pm6 \kms$
Mpc$^{-1}$ for Models 3--10, 3--15, and 3--20 respectively, while the
estimates based on M~100 (Mould et al. 1995) are $83$, $78$ and $74 \kms$
Mpc$^{-1}$ respectively.  The metal poor calibration (MLCS) moves to lower
distance and higher Hubble constant, while the metal rich calibration (M~100)
moves to higher distances and lower Hubble constants.  The comparison with
other distance indicators (Figure 2) is no worse than the other
cases, with the exception of NGC~5253.

\section{Conclusions}

We have systematically explored the problem of extragalactic 
Cepheid distances and their primary systematic errors.  While the 
details of some of the model implementations are certainly open to
criticism, we have tried to include or illustrate all the principal uncertainties.
When we implement the standard extragalactic
analysis method (Madore \& Freedman 1991) in Model 0, we find general agreement with existing
distances, with significant corrections only for the Type Ia
supernova calibration galaxies.  Our revised, distances to the SNIa 
MLCS (Riess et al. 1996) calibrating galaxies NGC~5253, NGC~4536,
and NGC~4639 produce a revised Hubble constant estimate of $69\pm8\kms$ Mpc$^{-1}$.
Our new distance estimates and uncertainties have the advantages of a homogeneous 
statistical treatment, a standard extinction model, and the inclusion of
the full uncertainties in the extinction and Cepheid model on the distances.   
In Model 0, the magnitude residuals for all galaxies are strongly correlated with the
extinction vector, and when we allow each Cepheid an individual extinction in Model 1, the 
typical magnitude residual drops from $0.29$ mag to $0.09$ mag.  The 
Galactic Cepheid community 
(e.g. Caldwell \& Coulson 1986, 1987, Caldwell \& Laney 1991, Fernie 1990, 
Fernie et al. 1995, Laney \& Stobie 1993, 1994, 1996) includes individual 
extinctions as a matter of routine.  Some extragalactic studies effectively
fit individual extinctions or use reddening free magnitudes (e.g. Tanvir et al.
1995, Freedman et al. 1990, 1991, 1992, Saha et al. 1996ab, 1997), but they
are usually not used in the final distance estimate.  As the sadly neglected
work of Gould (1994) emphasized, a correct statistical model must include
the effects of these correlations on the uncertainties in the distance
{\it independent of the physical interpretation for their origin.} The danger
of confusing extinction, temperature, and correlated systematic errors
affects only the interpretation of the model not the need to account for 
correlated residuals.  Including the correlations does not change the values for the 
distances and mean extinctions, but it does significantly reduce their uncertainties 
(see Figure 1).

In both of these models, many Cepheids require negative intrinsic extinctions
for the host galaxies, which is clearly unphysical.  The simple solution to
the problem is to raise the mean LMC extinction form the standard value of
$\langle E\rangle_{LMC}=0.1$ to $\gtorder 0.2$, as first suggested by 
Freedman et al. (1992) to solve the negative extinction estimates in NGC~300
and recently reintroduced by B\"ohm-Vitense (1997) in an analysis of the Galactic, 
LMC, SMC, and M31 Cepheids.   In our larger sample and after including the
scatter in the extinction, we find the problem is significantly worse and
that a mean LMC extinction of $\gtorder0.25$ would be required to eliminate
the problem completely.  However, even the high estimates of
the mean extinction in the LMC by Bessel (1991) based on polarization, HI 
column density, interstellar absorption lines, and intrinsic color estimates
correspond only to $\langle E\rangle_{LMC}\simeq 0.13$, and 90\% of the
LMC supergiants studied by Grieve \& Madore (1986) had extinctions less
than $0.18$, so simply raising the
mean LMC extinction is an implausible solution to the problem of negative
extinctions.  If we can reject raising the LMC extinction, then  
the need for negative extinctions is conclusive evidence for adding additional
physics to the Cepheid model or for substantial correlated systematic errors in 
the magnitudes beyond the simple HST calibration uncertainties.
 
The physical requirement that the extinction be positive
also means that the standard Cepheid distance estimates are systematically
biased by their equal treatment of positive and negative extinctions.  The
covariance of distance and extinction mean that low extinction estimates
are associated with higher distances, so any requirement for positivity in
the extinction will drive the distances to all Cepheid galaxies downwards.
The bias extends equally to galaxies with and without a negative extinction
problem because the extinction estimates and distances are also correlated
between galaxies -- if I force a reduction in the distance to one galaxy,
the correlations lead to a reduction in the distances to all galaxies.

A partial solution to the negative extinction problem is to use a finite
temperature distribution at fixed period so that blue Cepheids are interpreted 
as being hotter rather than having lower extinctions (see the discussion in
Freedman et al. 1992).  It is only a partial solution both because the instability
strip is narrow and because it cannot significantly change the mean extinction of the
distribution unless there are few Cepheids in the galaxy.
In Model 2 we allowed the Cepheids a temperature distribution,
and forced the positivity of the extinction for a range of LMC mean extinctions.
As expected from the bias in the standard distance estimates, the distance 
to every Cepheid galaxy decreased by mean values of $-0.20$, $-0.14$, and $-0.05$ mag
for LMC mean extinctions of $\langle E\rangle_{LMC}=0.10$, $0.15$, and $0.20$
respectively.  All Hubble constant estimates systematically rise,
although the effect is more dramatic for the low metallicity Type Ia Project galaxies
because of their bluer Cepheids.  In particular, the 
MLCS Hubble constant estimates become $80\pm6$, $78\pm6$ and $72\pm6\kms$ Mpc$^{-1}$
and the M100 (Mould et al. 1995) estimates become $90$, $88$, and $84\kms$ Mpc$^{-1}$
in order of increasing LMC extinction.  Model 2 is only a partial solution to
the negative extinction problem, and despite the addition of a temperature variable
for all 694 Cepheids, the likelihood of the solutions declined
significantly compared to Model 1.

The extinctions shows a rough correlation with the metallicity of the host galaxy,
as would be expected from theoretical models predicting that metal poor Cepheids should
be bluer than metal rich Cepheids (e.g. Stothers (1988), Stift (1990, 1995), Chiosi
et al. 1993).  In Model 3 we added a metallicity correction to the
Cepheid zero-point, similar to the model of Sasselov et al. (1996), and
we find a change in the mean V and I magnitude of $-0.14 \pm 0.14$ mag/dex
and a change in the V--I color of $0.13\pm0.04$ mag/dex.  The metallicity
terms change little with the assumptions about the LMC extinction.  
The trend with wavelength is that metal rich Cepheids become fainter
in U and B, start getting brighter at V, and show the largest increases
in the infrared, as expected from line-blanketing increasing
the opacity towards the blue and back-warming increasing the emission towards
the red.  While the quantitative estimates may be somewhat high, they
roughly agree with both the theoretical
expectations and other experimental or semi-empirical determinations 
(e.g. Caldwell \& Coulson 1986, 1987, Gieren et al. 1993, Stift 1995, 
Sasselov et al. 1996).  The change in color with metallicity is 
more important than the change in luminosity, so studies focusing only on 
the change in luminosity (Madore \& Freedman 1990, Gould 1994) miss the dominant effect.  
The  addition of the metallicity correction lets the model solve the negative 
extinction problems without simply driving 
the distance estimates downwards, and as discussed in Gould (1994) and Sasselov 
et al. (1996), it can also explain many of the discrepancies 
between low (Type Ia supernovae in low metallicity galaxies) and high 
(other distance indicators in high metallicity galaxies) estimates of the Hubble constant.  The MLCS
Hubble constant estimates from the low metallicity galaxies are $85\pm6$, $80\pm6$ and
$76\pm6\kms$ Mpc$^{-1}$, while the high metallicity M100 estimates are 
$83$, $78$ and $74\kms$ Mpc$^{-1}$ as we increase the mean LMC extinction.  
A metallicity dependence to the Cepheid distance scale is the most plausible
explanation of the negative extinction problem if it cannot be explained
by systematic errors in the photometry.  Otherwise, the qualitative effect 
is probably secure even if our quantitative
results are changed by improved data or metallicity estimates

There is no question that the models including temperature and metallicity are
beginning to push the limits of the data, although our inclusion of the 
covariances between variables should properly treat near degeneracies.  We 
can identify five areas for improvement in the data. The first is that the 
data on the LMC Cepheids needs to be expanded and observed in a more uniform
set of filters.  The enormous MACHO and EROS samples of Cepheids can almost
fill this role, but the calibration uncertainties need to be reduced, comparisons
to standard filters need to be better understood, and comparable data in 
additional filters is required. The second problem is that the 
HST Cepheid samples have only two-color photometry.  Since even the
simplest models that can fully probe the systematic uncertainties in
the Cepheid distance scale require extinction, temperature
and composition estimates for each individual Cepheid, a minimum
of four-color photometry is needed. The two additional filters should
be B (sensitive to extinction and metallicity) and 
H or K (sensitive to temperature and metallicity).  {\it Without additional
colors it will be difficult or impossible to cleanly address the
systematic problems in the Cepheid distance scale.}  The third problem
is that the formal uncertainties in the absolute calibration of the
HST magnitudes (about $0.05$ mag) are large compared to the effects
we are probing.  Some means of improving the absolute calibrations
is critical to better controlling the systematic uncertainties in 
the Cepheid distance scale.   The fourth problem is that many of the
Cepheid host galaxies lack metallicity measurements, forcing us to
rely on general correlations between galaxy type or luminosity and
metallicity.  

The final problem is that no matter how elaborate or rococo we make the analysis procedures for
Cepheid mean magnitudes, it is not the proper way to analyze the 
data.  The only ``correct'' way to analyze the Cepheid data is to directly
fit the observed light curves, in the spirit of the MLCS method for Type Ia supernovae 
(Riess et al. 1995).  Such an 
approach is critical to the extragalactic samples where heroic investments of
HST time still produce poor phase coverage in V and terrible phase coverage in I, let
alone adding data in the two additional filters needed to control 
systematic uncertainties.  Stetson (1996) has developed
a template method for identifying Cepheids and better estimating mean
magnitudes, and in our second paper on Cepheid distances (Kochanek 1997) we develop an improved
template method incorporating both color and velocity data.  Ideally, the
shapes of the light curves also constrain the gross physical properties
of the Cepheids and break some of the close degeneracies seen between 
temperature, extinction, distance, and metallicity that limit the validity
and accuracy of our current conclusions.

\acknowledgments Acknowledgements:
I would like to thank many people for material and intellectual help, particularly
J.A.R. Caldwell, N.R. Evans, W.L. Freedman, J. Huchra, M. Krockenberger and D. Sasselov.  
R. Kirshner, R. Kurucz, A. Saha and D. Welch
offered helpful comments.  J.A.R. Caldwell provided his latest compilation of phase-averaged 
data on the LMC and SMC Cepheids, and W.L. Freedman and A. Gould helped reconstruct the 
unpublished M31 data.

\clearpage

\def\hm{\hphantom{-}}
\begin{deluxetable}{lcrcrc}
\small
\tablenum{1}
\tablecaption{Cepheid Data}
\tablehead{ \colhead{Galaxy} &\colhead{Group} &\colhead{$N_i$} &\colhead{Bands} &\colhead{$E_f$} &\colhead{$[O/H]$} }
\startdata
Galaxy         &          &124    &UBVRIJHK     &--       &$\hm 0.30$ \nl
LMC            &          &71     &UBVRIJHK     &$ 0.063$ &$\hm 0.00$ \nl
SMC            &          &78     &UBVRIJHK     &$ 0.043$ &$-0.35$ \nl
M~33           &Local     &11     &BVRI    &$ 0.045$ &$0.58\pm0.04$ \nl
M~31           &Local     &30     &BVRI    &$ 0.080$ &$0.28\pm0.27$ \nl
NGC~300        &Sculptor  &13     &BVRI    &$ 0.025$ &$0.05\pm0.15$ \nl
M~81           &M~81      &31     &VI      &$ 0.038$ &$0.32\pm0.07$ \nl
M~101          &M~101     &29     &VI      &$-0.008$ &$0.02\pm0.00$ \nl
IC~4182        &          &19     &VI      &$-0.015$ &$-0.35$ \nl
NGC~5253       &Centaurus &8      &VI      &$ 0.048$ &$-0.25$ \nl
NGC~925        &NGC~1023  &73     &VI      &$ 0.065$ &$0.29\pm0.02$ \nl
M~96           &Leo~I     &7      &VI      &$ 0.015$ &$\hm 0.69$ \nl
NGC~3351       &Leo~I     &46     &VI      &$ 0.013$ &$0.94\pm0.03$ \nl
NGC~4536       &Virgo     &32     &VI      &$-0.005$ &$\hm 0.00$ \nl
M~100          &Virgo     &51     &VI      &$ 0.010$ &$0.84\pm0.04$ \nl
NGC~4496A      &Virgo     &56     &VI      &$ 0.003$ &$\hm 0.00$ \nl
NGC~4639       &Virgo     &14     &VI      &$ 0.013$ &$\hm 0.10$ \nl
\enddata
\tablecomments{ The metallicities are from Zaritsky et al. (1994), except for M~96 
   (Oey \& Kennicutt 1993) and M~31 (Blair et al. 1982).  Metallicities for the galaxies 
   not included in Zaritsky et al. (1994) were estimated from the metallicity-type or 
   metallicity-magnitude relations.  The metallicity uncertainties represent the rms
   range of metallicities assigned to the Cepheids in that galaxy based on the metallicity
   gradients if known. The foreground extinction estimates $E_f=E(B-V)_f$ are from Burstein \& Heiles (1984). }
\end{deluxetable}

\clearpage

\begin{figure}
\centerline{\psfig{figure=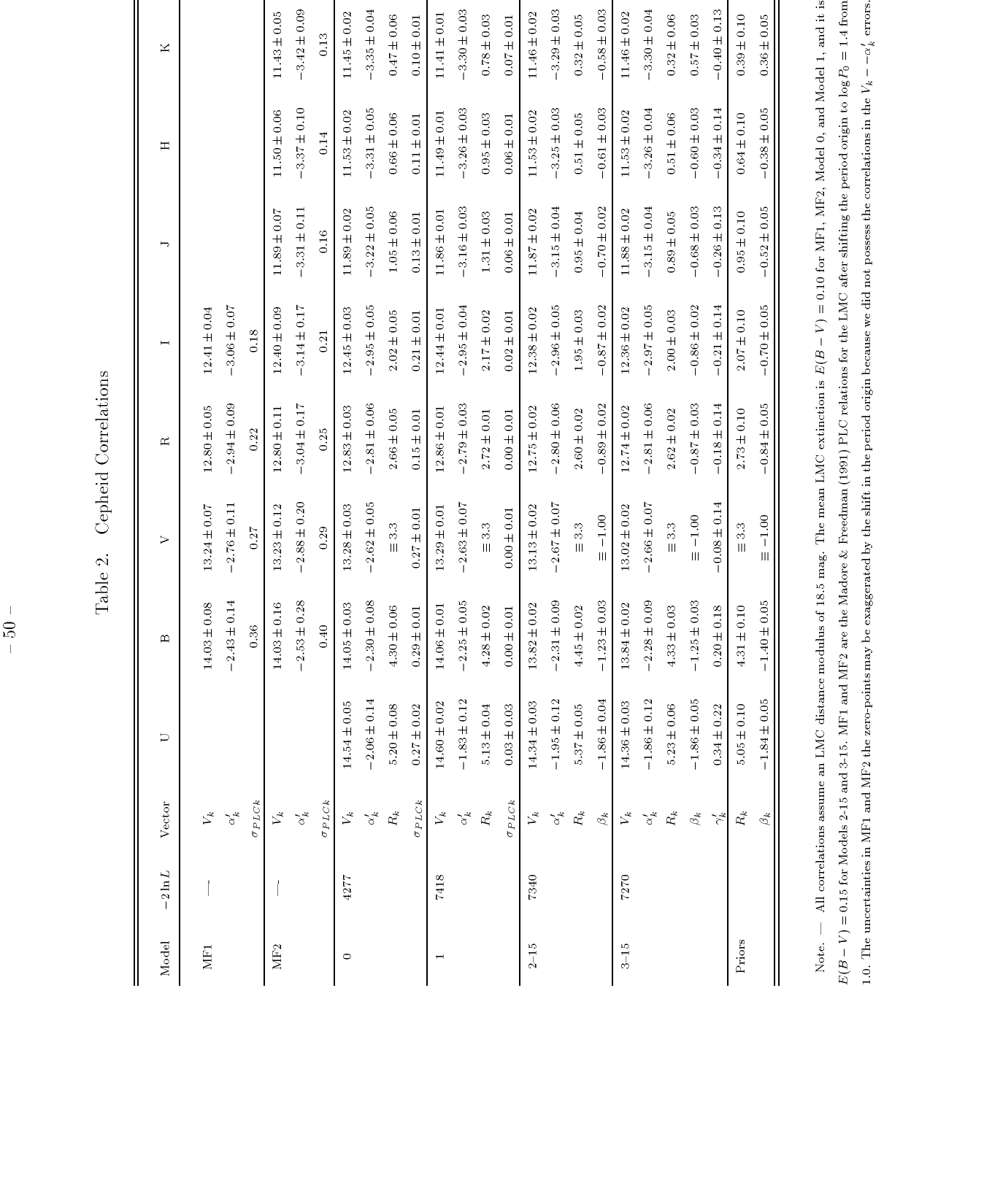,height=11.0in}}
\end{figure}

\clearpage

\begin{deluxetable}{rcrrrrr}
\scriptsize
\tablenum{3}
\tablecaption{Cepheid Distances \& Extinctions} 
\tablehead{ \colhead{Galaxy} &Var &\colhead{Published} &\colhead{Model 0} &\colhead{Model 1}  
     &\colhead{Model 2--15}  &\colhead{Model 3--15}  }
\startdata
Galaxy     
    &$R_0$                     &$14.52\pm0.15$ &$14.37\pm0.17$   &$14.12\pm0.11$   &$14.42\pm0.11$  &$14.53\pm0.12$ \nl
    &$\Theta_0(\kms)$          &$254\pm21$     &$244\pm18$       &$240\pm13$       &$241\pm14$      &$241\pm14$ \nl
LMC     
    &$\mu-\mu_{LMC}$           &               &$\equiv0.00$     &$\equiv0.0$      &$\equiv0.0$     &$\equiv0.0$ \nl
    &$\langle E(B-V) \rangle$  &$0.08$         &$\equiv0.10$     &$\equiv0.10$     &$\equiv0.15$    &$\equiv0.15$ \nl
    &$\sigma_E$                &$0.03$         &                 &$0.095\pm0.006$  &$0.079\pm0.007$ &$0.079\pm0.007$ \nl
SMC     
    &$\mu-\mu_{LMC}$           &$0.4\pm0.1$    &$0.47\pm0.02$    &$0.53\pm0.02$    &$0.48\pm0.02$   &$0.33\pm0.05$ \nl
    &$\langle E(B-V) \rangle$  &$0.04$         &$0.052\pm0.010$  &$0.037\pm0.003$  &$0.100\pm0.003$ &$0.137\pm0.014$ \nl
    &$\sigma_E$                &$0.02$         &                 &$0.082\pm0.004$  &$0.066\pm0.005$ &$0.066\pm0.005$ \nl
M~33     
    &$\mu-\mu_{LMC}$           &$6.14\pm0.09$  &$6.02\pm0.15$    &$6.00\pm0.06$    &$6.00\pm0.05$   &$6.35\pm0.09$ \nl
    &$\langle E(B-V) \rangle$  &$0.10\pm0.09$  &$0.161\pm0.050$  &$0.162\pm0.020$  &$0.209\pm0.016$ &$0.107\pm0.030$ \nl
    &$\sigma_E$                &               &                 &$0.055\pm0.013$  &$0.012\pm0.050$ &$0.010\pm0.050$ \nl
M~31     
    &$\mu-\mu_{LMC}$           &$5.94\pm0.14$  &$5.83\pm0.10$    &$5.77\pm0.04$    &$5.80\pm0.03$   &$5.97\pm0.05$ \nl
    &$\langle E(B-V) \rangle$  &$0.19\pm0.13$  &$0.194\pm0.032$  &$0.211\pm0.012$  &$0.249\pm0.010$ &$0.200\pm0.015$ \nl
    &$\sigma_E$                &               &                 &$0.106\pm0.012$  &$0.110\pm0.016$ &$0.105\pm0.019$  \nl
NGC~300
    &$\mu-\mu_{LMC}$           &$8.16\pm0.10$  &$8.11\pm0.14$    &$8.13\pm0.03$    &$8.02\pm0.03$   &dropped     \nl
    &$\langle E(B-V) \rangle$  &$-0.07\pm0.03$ &$-0.037\pm0.047$ &$-0.048\pm0.010$ &$0.034\pm0.008$ &            \nl
    &$\sigma_E$                &               &                 &$0.052\pm0.009$  &$0.055\pm0.016$ &            \nl
M~81
    &$\mu-\mu_{LMC}$           &$9.30\pm0.20$  &$9.22\pm0.21$    &$9.22\pm0.09$    &$9.05\pm0.09$   &$9.13\pm0.11$ \nl
    &$\langle E(B-V) \rangle$  &$0.03\pm0.05$  &$0.075\pm0.076$  &$0.071\pm0.027$  &$0.180\pm0.037$ &$0.183\pm0.036$ \nl
    &$\sigma_E$                &               &                 &$0.095\pm0.012$  &$0.072\pm0.018$ &$0.065\pm0.017$ \nl
M~101
    &$\mu-\mu_{LMC}$           &$10.84\pm0.17$ &$10.82\pm0.21$   &$10.84\pm0.11$   &$10.66\pm0.09$  &$10.56\pm0.11$ \nl
    &$\langle E(B-V) \rangle$  &$0.03\pm??$    &$0.015\pm0.079$  &$0.009\pm0.024$  &$0.123\pm0.033$ &$0.166\pm0.044$ \nl
    &$\sigma_E$                &               &                 &$0.095\pm0.010$  &$0.079\pm0.015$ &$0.079\pm0.017$ \nl
IC~4182
    &$\mu-\mu_{LMC}$           &$9.86\pm0.09$  &$ 9.98\pm0.23$   &$10.05\pm0.08$   &$9.81\pm0.11$   &$9.57\pm0.14$ \nl
    &$\langle E(B-V) \rangle$  &$-0.08\pm0.07$ &$-0.041\pm0.076$ &$-0.064\pm0.031$ &$0.065\pm0.039$ &$0.153\pm0.056$ \nl
    &$\sigma_E$                &               &                 &$0.081\pm0.013$  &$0.044\pm0.022$ &$0.044\pm0.023$  \nl
NGC~5253
    &$\mu-\mu_{LMC}$           &$9.58\pm0.10$  &$9.20\pm0.32$    &$9.23\pm0.16$    &$9.07\pm0.13$   &$8.90\pm0.15$ \nl
    &$\langle E(B-V) \rangle$  &$0.02\pm0.29$  &$0.071\pm0.125$  &$0.058\pm0.061$  &$0.165\pm0.046$ &$0.209\pm0.051$ \nl
    &$\sigma_E$                &               &                 &$0.080\pm0.023$  &$0.044\pm0.038$ &$0.054\pm0.032$ \nl
NGC~925
    &$\mu-\mu_{LMC}$           &$11.34\pm0.16$ &$11.33\pm0.18$   &$11.31\pm0.08$   &$11.18\pm0.10$  &$11.23\pm0.10$ \nl
    &$\langle E(B-V) \rangle$  &$0.13\pm0.08$  &$0.136\pm0.067$  &$0.137\pm0.035$  &$0.235\pm0.031$ &$0.244\pm0.034$ \nl
    &$\sigma_E$                &               &                 &$0.102\pm0.007$  &$0.069\pm0.010$ &$0.069\pm0.011$ \nl
M~96
    &$\mu-\mu_{LMC}$           &$11.82\pm0.16$ &$11.78\pm0.28$   &$11.83\pm0.10$   &$11.64\pm0.10$  &$11.92\pm0.15$ \nl
    &$\langle E(B-V) \rangle$  &$0.06\pm0.03$  &$0.061\pm0.117$  &$0.043\pm0.029$  &$0.159\pm0.043$ &$0.127\pm0.038$ \nl
    &$\sigma_E$                &               &                 &$0.093\pm0.023$  &$0.076\pm0.029$ &$0.067\pm0.030$ \nl
NGC~3351
    &$\mu-\mu_{LMC}$           &$11.51\pm0.19$ &$11.42\pm0.19$   &$11.41\pm0.06$   &$11.27\pm0.08$  &$11.59\pm0.14$ \nl
    &$\langle E(B-V) \rangle$  &$0.12\pm0.02$  &$0.122\pm0.073$  &$0.122\pm0.047$  &$0.222\pm0.033$ &$0.176\pm0.032$ \nl
    &$\sigma_E$                &               &                 &$0.122\pm0.013$  &$0.105\pm0.014$ &$0.104\pm0.014$ \nl
NGC~4536
    &$\mu-\mu_{LMC}$           &$12.60\pm0.13$ &$12.47\pm0.22$   &$12.35\pm0.11$   &$12.22\pm0.09$  &$12.11\pm0.12$ \nl
    &$\langle E(B-V) \rangle$  &$0.04\pm0.04$  &$0.077\pm0.079$  &$0.117\pm0.032$  &$0.215\pm0.035$ &$0.261\pm0.042$ \nl
    &$\sigma_E$                &               &                 &$0.083\pm0.011$  &$0.065\pm0.015$ &$0.064\pm0.015$ \nl
M~100 
    &$\mu-\mu_{LMC}$           &$12.54\pm0.17$ &$12.49\pm0.20$   &$12.50\pm0.10$   &$12.32\pm0.09$  &$12.60\pm0.15$ \nl
    &$\langle E(B-V) \rangle$  &$0.10\pm0.06$  &$0.075\pm0.072$  &$0.071\pm0.020$  &$0.182\pm0.031$ &$0.143\pm0.029$ \nl
    &$\sigma_E$                &               &                 &$0.117\pm0.011$  &$0.098\pm0.012$ &$0.097\pm0.012$ \nl
NGC~4496A 
    &$\mu-\mu_{LMC}$           &$12.53\pm0.14$ &$12.41\pm0.20$   &$12.43\pm0.11$   &$12.27\pm0.10$  &$12.17\pm0.13$ \nl
    &$\langle E(B-V) \rangle$  &$0.03\pm0.04$  &$0.061\pm0.074$  &$0.056\pm0.028$  &$0.162\pm0.037$ &$0.207\pm0.044$ \nl
    &$\sigma_E$                &               &                 &$0.094\pm0.008$  &$0.066\pm0.011$ &$0.066\pm0.011$ \nl
NGC~4639 
    &$\mu-\mu_{LMC}$           &$13.53\pm0.22$ &$13.61\pm0.32$   &$13.59\pm0.21$   &$13.24\pm0.11$  &$13.28\pm0.13$ \nl
    &$\langle E(B-V) \rangle$  &$??\pm??$      &$-0.083\pm0.114$ &$-0.078\pm0.072$ &$0.091\pm0.035$ &$0.093\pm0.039$ \nl
    &$\sigma_E$                &               &                 &$0.112\pm0.021$  &$0.097\pm0.030$ &$0.098\pm0.031$ \nl
\enddata
\tablecomments{
  The published extinction estimates for M~33, M~31, IC~4182, NGC~5253, NGC~4536, NGC~4496A, and NGC~4639 are not 
  directly comparable to our estimates because the authors used different extinction models. A ? indicates
  that no estimate was included in the published results. The value for the solar radius $R_0$ is the consensus
  value from Reid (1993), and the value for the circular velocity of the sun $\Theta_0$ 
  combines the $R_0$ estimate with the proper motion
  of Sgr A$^*$ (Backer 1996).  The relative distance of the LMC and the SMC and the extinction 
  values are from the review by Westerlund (1990), although Jacoby et al. (1990) find higher values of $0.13$ 
  for the LMC and $0.06$ for the SMC based on Balmer decrements in planetary nebulae.
  Our distances and the Westerlund (1990) value for the SMC are larger than the Caldwell \& 
  Laney (1991) values because of differences in defining the Cloud centers. {\bf The distance errors are
  very strongly correlated and cannot be treated as independent, random uncertainties!} }
\end{deluxetable}

\clearpage

\begin{deluxetable}{rcrrrrr}
\footnotesize
\tablenum{4}
\tablecaption{Comparison Distances}
\tablehead{ 
\colhead{Galaxy} &\colhead{EPM}             &\colhead{SNIa}           &\colhead{SBF}     &\colhead{SNIa/MLCS}  }
\startdata
                 &$\mu-18.5$                &$\mu-18.5$               &$\mu-\mu_{M31}$   &$\mu-\mu_{SN1981B}$ \nl
\hline
          M~31    &                          &                         &$\equiv0\pm0.06$  &   \nl
          M~81    &                          &                         &$3.26\pm0.08$     &   \nl
          M~101   &$10.85^{+0.28}_{-0.49}$   &                         &                  &   \nl
        IC~4182   &                          &$9.77^{+0.44}_{-0.55}$   &                  &   \nl
       NGC~5253   &$(9.56\pm0.2)$            &$9.51^{+0.30}_{-0.35}$   &$3.65\pm0.10$     &$-3.17\pm0.09$ \nl
       NGC~925    &$(11.63^{+0.35}_{-0.24})$ &                         &$5.53\pm0.09$     &               \nl
       M~96       &$(12.38^{+0.93}_{-1.37})$ &                         &$(5.76\pm0.06)$   &               \nl
       NGC~3351   &$(12.38^{+0.93}_{-1.37})$ &                         &$(5.76\pm0.06)$   &               \nl
       NGC~4536   &                          &$12.89^{+0.41}_{-0.51}$  &                  &$\equiv0\pm0.07$ \nl
       M~100      &$12.37^{+0.51}_{-0.67}$   &                         &                  &                \nl
       NGC~4639   &$13.00^{+0.48}_{-0.62}$   &                         &                  &$0.66\pm0.14$ \nl
\enddata
\tablecomments{ Absolute distances for the Cepheid galaxies based on the expanding photosphere method
  (EPM, Eastman, Schmidt \& Kirshner 1996) or physical models of Type Ia supernovae (SNIa, H\"oflich
  \& Khokhlov 1996), and relative distances based on the surface brightness fluctuation method (SBF, Tonry
  et al. 1996) and the MLCS method for Type Ia supernovae (SNIa/MLCS,
  Riess et al. 1996).  The SBF distances are relative to M31, and the MLCS distances
  are relative to SN~1981B in NGC~4536.  Values without parenthesis are distances to the Cepheid 
  galaxy, while values in parenthesis are for galaxies in the same group.
}  
\end{deluxetable}

\end{document}